\documentstyle[aps,preprint,tighten,epsf]{revtex}
\begin{document}
\preprint{\vbox{\hfill SHEP 96/27\\
                \null\hfill HLRZ 72/96\\
                \null\hfill WUB 96--39\\ }}

\title
{Ab Initio
Calculation of Relativistic Corrections to the Static Interquark
Potential I: SU(2) Gauge Theory}
\author{Gunnar S.\ Bali\thanks{Electronic mail: bali@hep.ph.soton.ac.uk}}
\address{Physics Department, The University of
Southampton\\Highfield, Southampton SO17 1BJ, England}
\author{Klaus Schilling\thanks{Electronic mail:
schillin@theorie.physik.uni-wuppertal.de}, Armin
Wachter\thanks{Electronic mail: wachter@hlrserv.hlrz.kfa-juelich.de}}
\address{HLRZ, Forschungszentrum J\"ulich, 52425 J\"ulich and\\
DESY, 22603 Hamburg, Germany\\ and\\
Fachbereich Physik, Bergische Universit\"at, Gesamthochschule
Wuppertal\\Gau\ss{}stra\ss{}e 20, 42097 Wuppertal, Germany}
\date{\today}\maketitle
\begin{abstract}
We test the capability of state-of-the-art lattice techniques for
a precise determination
of relativistic corrections to the static interquark
potential, by use of SU(2) gauge theory.  Emphasis is put on the
short range structure of the spin dependent potentials, with lattice
resolution $a$ ranging from $a\approx 0.04$~fm (at $\beta=2.74$) down to
$a\approx 0.02$~fm (at $\beta=2.96$) on volumes of $32^4$ and
$48^4$ lattice
sites. We find a new short range Coulomb-like contribution to the
spin-orbit potential $V_1'$.
\end{abstract}
\draft

\section{Introduction}
Quarkonia spectroscopy provides a wealth of information and thus
constitutes an important observational window to the phenomenology of
confining quark interactions.
It has been known for a long time that
purely phenomenological or QCD inspired potential models offer  a
suitable heuristic framework to understand the empirical  charmonium
($J/\psi$) and bottomonium ($\Upsilon$)
spectra~\cite{cornell,quigg,lucha}.

On a more fundamental level, one would prefer to start out from the
basic QCD Lagrangian to solve the heavy quarkonia bound state problem.
Two alternative strategies lend themselves for reaching this goal: (i)
direct extraction of the bound states on the lattice from an {\it
 effective nonrelativistic} lattice Lagrangian approximation
(NRQCD)~\cite{NRQCD}, (ii)
use of an effective nonrelativistic Hamiltonian framework through
the intermediary of potentials determined from lattice QCD.

Considerable efforts have been made recently to determine the
quarkonia spectra within the NRQCD approximation of QCD~\cite{nrqcd};
the notorious technical problems to determine excited states in the
Euclidean formulation have been tackled with remarkable success.  In
the alternative Schr\"odinger-Pauli setting, the technical problems are
shifted towards the lattice determination of $1/m^2$ corrections to
the potential.  The spin dependent (sd)~\cite{eich-fein,gromes}
and velocity
dependent (vd)~\cite{BBMP} contributions need to be extracted from
(Euclidian) time asymptotia of rather complex observables that
require renormalization and must obey constraints following
from Lorentz symmetry\cite{grome2,BBMP}.

First attempts to compute the relativistic corrections to the static
potential on the lattice have been pioneered in the mid
eighties~\cite{michael1,rebbi,forcrand,huntley}. In the meantime tremendous
improvements have been achieved both in computational power and
methods.  The central potential has been determined with high accuracy
in quenched QCD~\cite{pot,pot2,michael3} and, more recently, in full
QCD with two
dynamical flavors of light Wilson sea quarks~\cite{SESAM}. In view of
the general interest in the potential formulation of the meson binding
problem, renewed effort should be made to unravel the
structure of sd and vd potentials. This will provide us with a better
understanding of the structure of the interaction in the
intermediate distance
regime $0.15\ \mbox{fm} < r < 1\ \mbox{fm}$ which is of
tantamount
importance to the binding problem.

As a first step within this program we shall present in this paper a high
statistics study of the spin dependent forces in SU(2) gauge
theory. Though the two color formulation will not yet allow to proceed
to spectrum calculations we would expect the key features of
gluodynamical confinement to be revealed. In a follow-up
paper~\cite{SU3} (referred to as II) we shall apply our techniques
to the SU(3) case.

The present article is organized as follows: in Section~\ref{sec2}, we
provide an introduction into the Hamiltonian formulation of QCD
binding problems, the expected theoretical scenario of potentials as
well as the lattice observables from which to determine them.  In
Section~\ref{sec3}, useful lattice techniques will be
elaborated.  In particular we shall discuss the systematic
uncertainties of the approach.  The resulting SU(2) gauge potentials
will be presented in Section~\ref{sec4}.

\section{The heavy quark potential}
\label{sec2}
\subsection{Hamiltonian formulation of the meson binding problem}
Starting from a Foldy-Wouthuysen transformation of the Euclidean
quark propagator in an external gauge field, the asymptotic
($T\rightarrow\infty$) expression
$\langle W(R,T)\rangle\propto\exp(-\hat{V}_0(R)T)$
for expectation values of Wilson loops can be derived. 
$V_0(r)=a^{-1}\hat{V}_0(R)$ denotes the potential between static quarks,
separated by a distance
$r=aR$. $R$ and $T$ are the spatial and temporal extents
of the (rectangular) Wilson loop.
By perturbing the propagator in terms of the
inverse quark masses $m_1^{-1}$ and $m_2^{-1}$ around its static
solution, one arrives at the semi-relativistic Hamiltonian (in the CM
system, i.e.\ ${\mathbf p}={\mathbf p_1}=-{\mathbf p_2}$ and ${\mathbf L}=
{\mathbf L_1}={\mathbf L_2}$),
\begin{equation}
\label{ham}
H=\sum_{i=1}^2\left(m_i+\frac{p^2}{2m_i}-\frac{p^4}{8m_i^3}\right)
+V_0(r)+
V_{\mbox{\scriptsize sd}}(r,{\mathbf L},{\mathbf S_1},{\mathbf S_2})+
V_{\mbox{\scriptsize vd}}(r,{\mathbf p}),
\end{equation}
where the potential consists of a central part, sd~\cite{eich-fein,gromes} and
vd~\cite{BBMP} corrections\footnote{
Here, we just state the classical values for the couplings to the
potentials. Radiative corrections give rise to small deviations from
these tree level results~\cite{Pantaleone}. The corresponding one loop
coefficients have recently been determined
in the framework of heavy quark effective theory (HQET)
in Ref.~\cite{Chen}. In the unequal mass case additional contributions
appear, whose tree level coefficients are zero, that can also
be parametrized in
terms of $V_1'$ an $V_2'$.},
\begin{eqnarray}
V_{\mbox{\scriptsize sd}}(r,{\mathbf L},{\mathbf S_1},{\mathbf S_2})
&=&\left(\frac{{\mathbf L}{\mathbf S_1}}{m_1^2}
+ \frac{{\mathbf L}{\mathbf S_2}}{m_2^2}\right)
\frac{V_0'(r)+2V_1'(r)}{2r}\label{sdpo}\\
&+&\frac{{\mathbf L}({\mathbf S_1} + {\mathbf S_2})}{m_1m_2}
\frac{V_2'(r)}{r}
+\frac{S_1^iS_2^j}{m_1m_2}\left(R_{ij}V_3(r)+\frac{\delta_{ij}}{3}V_4(r)\right)
\nonumber
\end{eqnarray}
with
\begin{equation}
R_{ij}=\frac{r_ir_j}{r^2}-\frac{\delta_{ij}}{3}
\end{equation}
and
\begin{eqnarray}\label{vdpo}
V_{\mbox{\scriptsize vd}}(r,{\mathbf p})
&=&\frac{1}{8}\left(\frac{1}{m_1^2}+\frac{1}{m_2^2}\right)
\left(\nabla^2V_0(r)+\nabla^2V_a(r)\right)\\\nonumber
&-&\frac{1}{m_1m_2}\left\{p_i,p_j,S_{ij}\right\}_{\mbox{\scriptsize Weyl}}
+\sum_{k=1}^2\frac{1}{m_k^2}\left\{p_i,p_j,T_{ij}
\right\}_{\mbox{\scriptsize Weyl}}
\end{eqnarray}
with $S_{ij}=\delta_{ij}V_b(r)-R_{ij}V_c(r)$ and
$T_{ij}=\delta_{ij}V_d(r)-R_{ij}V_e(r)$.
The symbol $\{\cdot,\cdot,\cdot\}_{\mbox{\scriptsize Weyl}}=
\frac{1}{4}\{\cdot,\{\cdot,\cdot\}\}$
denotes Weyl ordering of the three arguments.
$V_1'$ -- $V_4$ are related to spin-orbit and spin-spin interactions,
$V_b$--$V_e$ to orbit-orbit interactions and the Darwin-like term
that incorporates $\nabla^2V_a$ modifies the
central potential\footnote{In following the convention of Ref.~\cite{BBMP}
we have included this term into $V_{\mbox{\scriptsize vd}}$ although
it does not
explicitly depend on the velocity.}.
$V_1'$ -- $V_4$ and $\nabla^2V_a$ --
$V_e$ can be computed from lattice correlation functions
in Euclidean time of Wilson loop like operators.
Pairs of the potentials are related by Lorentz invariance
to the central
potential~\cite{grome2,BBMP}:
\begin{eqnarray}
V_2'(r)-V_1'(r)&=&V_0'(r),\label{grom}\\
V_b(r)+2V_d(r)&=&\frac{r}{6}V_0'(r)-\frac{1}{2}V_0(r),\\
V_c(r)+2V_e(r)&=&-\frac{r}{2}V_0'(r)\label{bram},
\end{eqnarray}
such that only six sd and vd potentials turn out to be truly
independent.

In the present SU(2) investigation, we restrict ourselves to the spin
dependent terms. Note, that the Hamiltonian Eq.~(\ref{ham}) contains
dimension six operators and thus is not renormalizable. For this
reason the theory --- being truncated at order $1/m^2$ --- is only an
effective one with validity range of small gluon momenta (compared to
the heavy quark masses).  This very fact gave rise to a discrepancy
between the Eichten-Feinberg-Gromes formulae Eq.~(\ref{sdpo}) and
perturbative expansions~\cite{radford,Pantaleone} in powers of the coupling,
$g$, where additional logarithmic mass dependencies occurred from
dimensional regularization. The underlying problem is now solved and
one loop matching coefficients between the effective Hamiltonian
and QCD have been obtained from HQET~\cite{Chen}.

\subsection{Expectations on sd potentials}
\label{expect}
In addition to the exact constraint, Eq.~(\ref{grom}), derived by
Gromes~\cite{grome2}, some approximate relations between the sd
potentials are anticipated from exchange symmetry arguments.
We start from the standard assumption that the origin of the
central potential is due to vector and scalar-like gluon exchange
contributions.
Given the fact that a vector-like exchange can at most
grow logarithmically with $r$~\cite{grome}, the nature of the
linear part of the
confining potential can only be scalar. As we will see, $V_2'(r)$ is
short ranged, such that the confining part only contributes to
$V_1'(r)$. This leads us to expect $V_2'(r)$ to be purely vector-like.
Under the additional assumptions that pseudoscalar contributions can
be neglected and that $V_1'$ does not contain a vector-like
contributions, one ends up with the scenario of interrelations~\cite{grome2},
\begin{eqnarray}
\label{exv31}
V_3(r)&=&\frac{V_2'(r)}{r}-V_2''(r),\\\label{exv3}
V_4(r)&=&2\nabla^2V_2(r),
\end{eqnarray}
which of course has to be in agreement with leading order perturbation
theory. However, Eqs.~(\ref{exv31})--(\ref{exv3}) hold true for any
effective gluon propagator that transforms like a Lorentz vector.
In principle, to this order in $m^{-1}$, $V_3$ and $V_4$ could contain
additional pseudoscalar pieces.

Tree-level continuum perturbation theory (see Appendix) yields the following
expectations for the central, spin-orbit and spin-spin potentials:
\begin{eqnarray}
\label{pertu}
V_0(r)&=&-\frac{e}{r}\label{pertu0},\\\label{pertu1}
V_1'(r)&=&0,\\\label{pertu2}
V_2'(r)&=&\frac{e}{r^2},\\\label{pertu3}
V_3(r)&=&\frac{3e}{r^3},\\\label{pertu4}
V_4(r)&=&8\pi e\delta^3(r),
\end{eqnarray}
where $e=C_F\alpha_s$ and $C_F=3/4$ for SU(2). Note, that
combining the perturbative result on $V_2'(r)$ with the
Gromes relation, Eq.~(\ref{grom}), and a funnel type parametrization of
the central potential~\cite{cornell},
\begin{equation}
V_0(r)=-\frac{e}{r}+\kappa r,
\end{equation}
the first spin-orbit potential should take the form
\begin{equation}
\label{v1ex}
V_1'(r)=-\kappa.
\end{equation}

In the Appendix we shall also derive the corresponding tree level
results for the
lattice potentials. We will show that an exact lattice relation
between $\tilde{V}_1'$ and $\tilde{V}_2'$ and a linear combination of
the central
potential, taken at different lattice coordinates, in analogy to the
Gromes relation does not exist. However, the relation should be
retrieved in the continuum limit
and approximately hold within the scaling
region on the lattice.

\subsection{How to compute the potentials}
In the potential picture, the quarks interact instantaneously
through a potential that only
depends on the distance, spins, and velocities of the sources,
Eqs.~(\ref{ham}), (\ref{sdpo}) and (\ref{vdpo}).
All time dependence has been separated and implicitly included 
into coefficient functions of various interaction terms, the so called
sd and vd potentials.
These can be computed by a non-perturbative integration
over gluonic interactions. Therefore, the sd and vd
potentials incorporate a summation over all possible
interaction times, $t$. One obtains the following expressions
in terms of expectation values in presence of a gauge
field background for the sd
potentials~\cite{eich-fein,gromes}\footnote{We have put the expressions
into a form that is more suitable for lattice simulations.
Via spectral decompositions of the underlying correlation
functions, equality of the expressions to those
of Refs.~\cite{eich-fein,gromes}
can easily be shown.}:

\begin{eqnarray}
\label{ef_1}
\frac{R_k}{R}\tilde{V}_1'({\mathbf R}) &=& 
2\epsilon_{ijk}
\lim\limits_{\tau\to\infty}
\int_0^\tau 
\!dt\,t\, \langle\langle \hat{B}_i({\mathbf 0},0)\hat{E}_j({\mathbf
0},t)
\rangle\rangle_W, \\
\label{ef_2}
\frac{R_k}{R}\tilde{V}_2'({\mathbf R}) &=& \epsilon_{ijk}
\lim\limits_{\tau\to\infty}
\int_0^\tau
\!dt\, t\, \langle\langle \hat{B}_i({\mathbf
0},0)\hat{E}_j({\mathbf R},t)
\rangle\rangle_W, \\
\label{ef_3}
R_{ij}\tilde{V}_3({\mathbf R}) &=& 2\lim\limits_{\tau\to\infty}
\int_0^\tau 
\!dt\,\left[\langle\langle \hat{B}_i({\mathbf 0},0)
\hat{B}_j({\mathbf R},t)\rangle\rangle_W\right.\\\nonumber
&-&\frac{\delta_{ij}}{3}\left.
\langle\langle {\mathbf \hat{B}}({\mathbf 0},0){\mathbf
\hat{B}}({\mathbf R},t)
\rangle\rangle_W\right],\\
\label{ef_4}
\tilde{V}_4({\mathbf R}) &=& 2\lim\limits_{\tau\to\infty}
\int_0^\tau 
\!dt\, \langle\langle {\mathbf \hat{B}}({\mathbf 0},0)
{\mathbf \hat{B}}({\mathbf R},t)\rangle\rangle_W.
\end{eqnarray}
With $a\rightarrow 0$, the above potentials should approach their continuum
counterparts and rotational invariance should be restored,
$\tilde{V}_1'({\mathbf R})=a^{2}V_1'(aR)$,
$\tilde{V}_2'({\mathbf R})=a^{2}V_2'(aR)$,
$\tilde{V}_3({\mathbf R})=a^{3}V_3(aR)$ and
$\tilde{V}_4({\mathbf R})=a^{3}V_4(aR)$.

The operator $\langle\langle F_1F_2\rangle\rangle_W$ is 
defined as follows:
\begin{equation}
\label{dex}
\left\langle\langle F_1F_2\rangle\right\rangle_W=
\frac{\langle\mbox{Tr}\,{\mathcal P}
\left[\exp\left(ig\int_{\partial W} dx_\mu\,A_\mu\right)F_1F_2\right]\rangle}{
\langle\mbox{Tr}\,{\mathcal P}
\left[\exp\left(ig\int_{\partial W} dx_\mu\,A_\mu\right)\right]\rangle},
\end{equation}
where $\partial W$ denotes a closed path (the contour of a Wilson loop
$W({\mathbf R},T)$).
${\mathcal P}$ denotes path ordering of the arguments. Following
Ref.~\cite{huntley} we have chosen the following discretized version
of Eq.~(\ref{dex}) for the case of two color field insertions:
\begin{equation}
\label{defcor}
\left\langle\langle
\hat{F}_1\hat{F}_2\rangle\right\rangle_W
=-\frac{\left\langle
{\mathcal P}\left[W\,(\Pi_1-\Pi^{\dagger}_1)(\Pi_2-\Pi^{\dagger}_2)\right]
\right\rangle\left\langle\left. W\right.\right\rangle}
{\left\langle {\mathcal
P}\left[W\,(\Pi_1+\Pi^{\dagger}_1)\right]\right\rangle
\left\langle {\mathcal
P}\left[W\,(\Pi_2+\Pi^{\dagger}_2)\right]\right\rangle},
\end{equation}
where the subscripts
$1$ and $2$ represent the multi-indices $(n_1,\mu_1,\nu_1)$ and
$(n_2,\mu_2,\nu_2)$, respectively. $n_i$ are integer valued
four-vectors. The above ratio of ``eared'' Wilson
loops is visualized in Fig.~\ref{mren}.
In order to avoid imaginary phases and
factors $g^2a^4$ from Eqs.~(\ref{ef_1})--(\ref{ef_4}),
we use the following conventions for the electric and magnetic fields:
\begin{equation}
\hat{F}_{\mu\nu}=ga^2F_{\mu\nu}\quad,\quad
\hat{E}_i= \hat{F}_{i4}\quad,\quad
\hat{B}_i=\frac{1}{2}\epsilon_{ijk}\hat{F}_{jk}.
\end{equation}

In what follows, we have chosen $\Pi_{\mu\nu}(n)$ to be the average of the
four plaquettes, enclosing the lattice point
$n$,
\begin{equation}\label{epi}
\Pi_{\mu\nu}(n) = \frac{1}{4}\left[P_{\mu,\nu}(n)+ P_{\mu,\nu}(n)
+P_{-\mu,-\nu}(n)+P_{-\mu,\nu}(n)\right]
\end{equation}
with
\begin{equation}
P_{\mu,\nu}(n)=U_{\mu}(n)U_{\nu}(n+\hat{\mu})U^{\dagger}_{\mu}(n+\hat{\nu})
U^{\dagger}_{\nu}(n)\quad,\quad
U_{-\mu}(n)=U^{\dagger}_{\mu}(n-\hat{\mu}).
\end{equation}
This choice of $\Pi$ makes Eq.~(\ref{defcor}) correct up to
order $a^2$, the discretization error of the Wilson action, used
for generating the gauge field background.

In practical computation, the temporal extent $T$ of the Wilson loop
$W$ within Eqs.~(\ref{ef_1})--(\ref{ef_4}) is adapted according to the
formula $T=t+2\Delta t$.  We choose to keep the minimal distance,
$\Delta t$, between the ``ears'' and the spatial closures of the
Wilson loop fixed.  Note, that, strictly speaking,
Eqs.~(\ref{ef_1})--(\ref{ef_4}) apply for the limit $\Delta
t\rightarrow\infty$.  Our coordinates are such that the Wilson loop
$W({\mathbf R},t+2\Delta t)$ extends from $n_4=-\Delta t$ to
$n_4=t+\Delta t$ into the temporal direction and from ${\mathbf
  n}={\mathbf 0}$ to ${\mathbf n}={\mathbf R}$ into the spatial
directions.  Physically speaking, $\Delta t$ represents the time we
allow the gluon field to decay into the ground state, after (before)
creation (annihilation) of the $q\bar{q}$-state. Hence this
deexcitation time must be considered as an important control parameter
of our measurements.

\section{Lattice simulations}
\label{sec3}
\subsection{Simulation parameters}
\label{sec31}
Since, apart from $V_1'$, all sd potentials are expected to be
short ranged, we will focus our attention onto short distance
properties, i.e.\ aim  at the smallest lattice resolution possible
on our computers. The present simulations have been performed on
$V=L_{\sigma}^3L_{\tau}=16^4$, $32^4$ and
$48^4$ lattices at $\beta=2.74$ and $\beta=2.96$
which correspond to lattice spacings $a\approx 0.041$~fm and
$a\approx 0.022$~fm, respectively (Table~\ref{et1}). The number of
independent Monte Carlo configurations $n_{\mbox{\scriptsize conf}}$,
generated at each set of parameters, is included into the Table. 
The above physical scales have been adjusted such that the string tension
comes out to be\footnote{This scale can
only provide a rough orientation as here we are simulating
SU(2) gauge theory, not full QCD.} $\sqrt{\kappa}=440$~MeV. Using such small
physical volumes, finite size effects (FSEs) have to be investigated. This
will be done by comparing results obtained on a $16^4$ lattice with
$32^4$ results at $\beta=2.74$.
Finite lattice resolution (i.e.\ finite $a$)
effects are investigated by relating results, obtained at the two
different values of the coupling.

\subsection{Updating algorithm}
The numerical calculations are performed on lattices with hypercubic
geometry and periodic boundary conditions in all four directions.
Throughout the simulation the standard Wilson action
\begin{equation}
S_W=-\beta\sum_{n,\mu>\nu}\frac{1}{2}\mbox{Tr} P_{\mu,\nu}(n)
\end{equation}
with $\beta=4/g^2$ has been used.

For the updating of the gauge fields, a hybrid of heatbath and
overrelaxation algorithms has been implemented~\cite{hor}. The 
Fabricius-Haan heatbath sweeps~\cite{fh}
have been randomly mixed with the overrelaxation step with probability
$1/14$. The links have been visited in lexicographical ordering within
hypercubes of $2^4$ lattice sites, i.e.\ within each such hypercube,
first all links pointing
into direction $\hat{1}$ are visited site by
site, then all links in direction $\hat{2}$ etc.. After at least 2000
heatbath thermalization sweeps, measurements are taken every 100
or 200 sweeps at $\beta=2.74$ and $\beta=2.96$, respectively, to
ensure decorrelation. We find no signs of any autocorrelation effects
between successive configurations
within any of the measured observables.

\subsection{Link integration}
Statistical fluctuations have been reduced
by ``integrating out'' temporal links that appear within the Wilson
loops and the electric ears analytically, wherever possible.
By ``link integration'' we mean the following substitution~\cite{parisi}:
\begin{equation}
U_4(n)\longrightarrow W_4(n)=\frac{\int_{SU(2)}\!dU\,Ue^{-\beta
S_{n,4}(U)}}
{\int_{SU(2)}\!dU\,e^{-\beta S_{n,4}(U)}}
\end{equation}
with
\begin{equation}
S_{n,\mu}(U)=-\frac{1}{2}\mbox{Tr}\left(UF_{\mu}^{\dagger}(n)\right),
\end{equation}
and
\begin{equation}
\quad F_{\mu}(n)=\sum_{\nu\neq\mu}U_{\nu}(n)U_{\mu}(n+\hat{\nu})
U_{\nu}^{\dagger}(n+\hat{\mu}).
\end{equation}
$W_4(n)$ is in general not an SU(2) element anymore.

In this way, time-like links are replaced
by the mean value they take in the neighborhood of
the enclosing staples $F_4(n)$. 
Only those links that do not share a common plaquette can be
integrated independently.

In case of SU(2) gauge theory, $W_4(n)$ can be calculated analytically,
\begin{equation}
W_4(n)=\frac{I_2(\beta f_{\mu}(n))}{f_{\mu}(n)I_1(\beta
f_{\mu}(n))}F_{\mu}(n),
\end{equation}
where $f_{\mu}(n)=\sqrt{\det(F_{\mu}(n))}$.
$I_n$ denote the modified Bessel functions.

\subsection{Smearing}
In order to achieve a satisfying overlap between the 
quark-antiquark ground state and the state created
by the spatial parts of the Wilson loop,
we have applied a smearing procedure~\cite{schlicht,APE}
by iteratively replacing each
spatial link
\(U_i(n)\) within the Wilson loop, by a ``fat'' link,
\begin{equation}
\label{sme}
U_i(n)\rightarrow N\left(\alpha\, U_i(n)+\sum_{j\neq
i}U_j(n)U_i(n+\hat{j})U_j^{\dagger}(n+\hat{i})\right),
\end{equation}
with the appropriate normalization $N$
and free parameter $\alpha$.
We find satisfactory ground state enhancement with the parameter choice
$n_{\mbox{\scriptsize iter}}=150$ and $\alpha=2$.
In Fig.~\ref{v0} the resulting central interquark potentials are
displayed. In Fig.~\ref{overlaps}, the corresponding ground state
overlaps $C_0(R)$ 
are shown as a function of the source separation at $\beta=2.74$.
As can be seen, all overlaps are well above 0.9.

\subsection{Spectral decomposition}
In this section we will discuss the control of excited state
contributions at finite deexcitation time $\Delta t$. As explained
above, the spatial transporters within the Wilson loops have been
smeared to suppress such pollutions from the very beginning, allowing
to work with moderate values of $\Delta t$. This is vitally important,
as statistical errors increase with the size of the Wilson loop. In
practice, we have decided to keep $\Delta t$ fixed at a value suited
for sufficient deexcitation and to increase the temporal extent of the
Wilson loop according to $T=t+2\Delta t$, i.e.\ with the
separation $t$ between the two ears. We found
$\Delta t = 2$ to be appropriate.

Previous authors~\cite{rebbi,forcrand} have replaced the integrals
over interaction
times by discrete sums.
This results in cut-off errors due to the finiteness of $\tau$ as well
as additional order $a^2$ integration errors. Both sources
of systematic uncertainties can be
significantly reduced by exploiting transfer matrix techniques.
For illustration of such techniques we start from the static potential
which can be computed from Wilson loops:
at Euclidean time $t=0$, a creation operator,
\begin{equation}
\label{create}
\Gamma^{\dagger}_{\mathbf R}=q({\mathbf 0})U({\mathbf 0}\rightarrow {\mathbf R})q^{\dagger}({\mathbf R})
\end{equation}
with a gauge
covariant transporter
$U({\mathbf 0}\rightarrow {\mathbf R})$ is applied to the vacuum
state $|0\rangle$. $q^{\dagger}({\mathbf R})$ creates a heavy quark
spinor at position
${\mathbf R}$. The $q\bar{q}$ pair is then propagated
to $t=T$ by static
Wilson lines in presence of the gauge field background, and finally
annihilated by $\Gamma_{\mathbf R}$. A spectral decomposition of the
Wilson loop exhibits the following behavior (${\mathcal T}=e^{-aH}$
denotes the
transfer matrix, ${\mathcal T}|n\rangle=e^{-aE_n}|n\rangle$):

\begin{eqnarray}
P_W=
\langle W({\mathbf R},T)\rangle&=&
\frac{\mbox{Tr}\left(\Gamma_{\mathbf R}{\mathcal T}^T\Gamma_{\mathbf R}^{\dagger}{\mathcal T}^{L_{\tau}-T}\right)}
     {\mbox{Tr}\left({\mathcal T}^{L_{\tau}}\right)}\nonumber\\\label{expa1}
&=&\frac{1}{\sum_m e^{-\hat{E}_mL_{\tau}}}\sum_{m,n}\left|\langle
m|\Gamma_{\mathbf R}|n,{\mathbf R}\rangle\right|^2
e^{-\hat{V}_n({\mathbf R})T}e^{-\hat{E}_m(L_{\tau}-T)}\\\nonumber
&=&
\sum_n |d_n({\mathbf R})|^2
e^{-\hat{V}_n({\mathbf R})T}\times\left(1+{\mathcal O}
\left(e^{-E_1(L_{\tau}-T)}\right)\right)
\end{eqnarray}
with 
\begin{equation}
d_n({\mathbf R})=\langle 0|\Gamma_{\mathbf R}|n,{\mathbf R}\rangle\quad,\quad
C_n({\mathbf R})=|d_n({\mathbf R})|^2.
\end{equation}
$|n,{\mathbf R}\rangle$ is the
$n$th eigenstate of a $q\bar{q}$ pair, separated by a distance
${\mathbf R}$. $|d_n({\mathbf R})|^2$ takes a positive value,
whenever such a state has a
non-vanishing overlap to the creation operator
$\Gamma_{\mathbf R}^{\dagger}$, applied to the vacuum.
$|n\rangle$ is the $n$th pure glue eigenstate (glueball).
$\hat{V}_n({\mathbf R})$ denotes the $n$th excitation of the $q\bar{q}$
potential and the vacuum energy $E_0$ has been set to zero. $E_1$ is the mass
gap, i.e.\ the mass of the $A_1^+$ glueball. Due to its heavy mass
and $L_{\tau}\gg T$, such back-propagating terms can be neglected to
high accuracy. The overlaps are normalized,
$\sum_n C_n({\mathbf R}) = 1$, such
that a ground state overlap $C_0({\mathbf R})\approx 1$
implies $C_n({\mathbf R})\ll 1$ for all $n>0$. $C_0({\mathbf R})$ can be
increased by optimizing the path combination
$U({\mathbf 0}\rightarrow {\mathbf R})$ within the
creation operator Eq.~(\ref{create}) (smearing).

Now, let us define the operators,
\begin{equation}
{\mathcal F}_i=q\frac{1}{2i}
\left(\Pi_i-\Pi_i^{\dagger}\right)q^{\dagger}\quad,\quad
{\mathcal G}_i=q\frac{1}{2}\left(\Pi_i+\Pi_i^{\dagger}\right)q^{\dagger},
\end{equation}
i.e.\ $q\Pi_iq^{\dagger}={\mathcal G}_i+i{\mathcal F}_i$,
for the chromo field insertions within the nominator and denominator of
Eq.~(\ref{defcor}). $i = ({\mathbf n}_i,\mu_i,\nu_i)$ denotes the position of
the insertion as well as the color field component. In the present
case, ${\mathbf n}_i$ either takes the position
${\mathbf 0}$ or ${\mathbf R}$. Let
\begin{equation}\label{gm}
f_{mn}^i({\mathbf R}) = \langle m,{\mathbf R} | {\mathcal F}_i | n,{\mathbf R}\rangle\quad,\quad
g_{mn}^i({\mathbf R}) = \langle m,{\mathbf R} | {\mathcal G}_i | n,{\mathbf R}\rangle.
\end{equation}
Hermiticity of ${\mathcal F}_i$ and ${\mathcal G}_i$ implies
$f^{i*}_{mn}=f^i_{nm}$ and
$g^{i*}_{mn}=g^i_{nm}$.

A spectral decomposition of a Wilson loop, $W({\mathbf R},T)$ with
ears ${\mathcal F}_1$ and ${\mathcal F}_2$ inserted 
at times $n_4=0$ and $n_4=t$,
respectively (The Wilson loop extends from $n_4=-\Delta t$ to
$n_4=t+\Delta t$.), yields (where we have neglected the
back-propagation of pure glue states):
\begin{eqnarray}
P_W^{12}&=&-\frac{1}{4}\left\langle
{\mathcal P}\left[W({\mathbf R},T)\left(\Pi_1-\Pi_1^{\dagger}\right)
\left(\Pi_2-\Pi_2^{\dagger}\right)\right]\right\rangle\nonumber\\
&=&
\frac{\mbox{Tr}\left(\Gamma{\mathcal T}^{\Delta t}{\mathcal F}_1{\mathcal T}^{t}
{\mathcal F}_2{\mathcal T}^{\Delta t}\Gamma^{\dagger}{\mathcal T}^{L_{\tau}-T}\right)}
{\mbox{Tr}\left({\mathcal T}^{L_{\tau}}\right)}\\\nonumber
&=&e^{-\hat{V}_0T}\sum_{l,m,n}\mbox{Re}\left(
d_ld_n^*f_{lm}^1f_{mn}^2\right)
e^{-(\Delta \hat{V}_l+\Delta \hat{V}_n)\Delta t} e^{-\Delta \hat{V}_mt}
\end{eqnarray}
with $\Delta \hat{V}_n=\hat{V}_n-\hat{V}_0$. For better readability,
the ${\mathbf R}$
dependence of $\Gamma_{\mathbf R}$, $\hat{V}_n({\mathbf R})$, $d_n({\mathbf
R})$ as well as $f_{mn}^i({\mathbf R})$ has been omited from the above
equation.

Next, let us investigate the behavior of the Wilson loops with one
ear, appearing in the denominator of Eq.~(\ref{defcor}). For an insertion of
${\mathcal G}_i$ at time
$n_4=0$ or $n_4=t$ we find,
\begin{eqnarray}
Q^i_W&=&
\frac{1}{2}
\left\langle{\mathcal P}\left[W({\mathbf R},T)
\left(\Pi_i+\Pi_i^{\dagger}\right)\right]\right\rangle\nonumber\\
&=&
\frac{\mbox{Tr}\left(\Gamma{\mathcal T}^{\Delta t}{\mathcal G}_i{\mathcal
T}^{t+\Delta t}\Gamma^{\dagger}{\mathcal T}^{L_{\tau}-T}\right)}
{\mbox{Tr}\left({\mathcal T}^{L_{\tau}}\right)}\\\nonumber
&=&e^{-\hat{V}_0T}\sum_{m,l}\mbox{Re}\left(d_ld_m^*g_{lm}^i\right)
e^{-(\Delta \hat{V}_l+\Delta \hat{V}_m)\Delta t} e^{-\Delta \hat{V}_mt}.
\end{eqnarray}

In combining the above expressions we finally obtain ,
\begin{equation}\label{eq:correl}
\left\langle\left\langle
\hat{F}_1\hat{F}_2\right\rangle\right\rangle_W
=\frac{P_W^{12}P_W}{Q_W^1Q_W^2}=
\sum_m D^{12}_me^{-\Delta \hat{V}_mt}\times\left(1+E^{12}_me^{-\Delta
\hat{V}_1\Delta t}+\cdots \right)
\end{equation}
with
\begin{equation}
\label{dm}
D^{12}_m = \frac{\mbox{Re}\left(f^1_{0m}
f^2_{m0}\right)}{g^1_{00}
g^2_{00}}
\end{equation}
and
\begin{equation}
\label{em}
E^{12}_m = \frac{\mbox{Re}\left[(d_1/d_0)\left(f_{1m}^1f_{m0}^2+
f_{0m}^1f_{m1}^2\right)\right]}
{\mbox{Re}\left(f_{0m}^1f_{m0}^2\right)}
-\frac{\mbox{Re}\left[(d_1/d_0)g^1_{10}\right]}{g^1_{00}}
-\frac{\mbox{Re}\left[(d_1/d_0)g^2_{10}\right]}{g^2_{00}}.
\end{equation}
Again, all constants are understood to depend on ${\mathbf R}$.

Note, that unwanted excited state contributions are suppressed by
the ratio $d_1/d_0$ as well as by $e^{-\Delta \hat{V}_1\Delta t}$.
The smallest value of $\Delta t$ that appears within an integral over
interaction times will determine the reliability of the
result. The bosonic string picture expectation is
$\Delta \hat{V}_1(R)= \pi/R$~\cite{luescher}. This has been
qualitatively confirmed in numerical studies~\cite{huntley2,green},
such that $\Delta t =2$ yields an excited state suppression by a
factor $\exp(-4\pi/R)$ for large $R$, in addition to the ratio
$d_1/d_0$.

The creation operator $\Gamma^{\dagger}$ projects only onto states
within the $A_{1g}$ representation of the appropriate symmetry group
$D_{4h}$~\cite{huntley2}. The lowest continuum angular momentum to
which it couples is $L=0$. The hybrid ($L=1$) state $E_u$ is the next
excitation. The combination of magnetic ears we use, applied to a
$q\bar{q}$ state within the $A_{1g}$ representation, results in a pure
$E_u$ state which has no overlap with $A_{1g}$, such that
$D_0^{12}=0$ within all correlation functions of interest, i.e.\ all
correlators decay exponentially with Euclidean time $t$. This does not
hold true for some of the vd potentials.

Unlike in the case of the central potential, where the sum of the
(non-negative) overlap coefficients $C_m$ is normalized to one,
the $D_m^{12}$ are not normalized and can be negative. However, due to
invariance 
under time inversion, $\sum_m D_m^{12}=0$ in case of $V_1'$ and $V_2'$
since the correlation function has to vanish at $t=0$.
In combining Eq.~(\ref{eq:correl}) with 
Eqs.~(\ref{ef_3}) or
(\ref{ef_4}), we obtain for $\tilde{V}_3$ or $\tilde{V}_4$:
\begin{equation}
\tilde{V}_{3,4} \propto \sum_{m>0}\int_0^{\infty}\!dt\,D_m^{12}
e^{-\Delta\hat{V}_mt}=
\sum_{m>0} \frac{D_m^{12}}{\Delta\hat{V}_m}
\end{equation}
with appropriate color field positions ${\mathbf
n}_1={\mathbf 0}, {\mathbf n}_2={\mathbf R}$ and components
$\mu_1, \nu_1, \mu_2, \nu_2$.
Eqs.~(\ref{ef_1}) and (\ref{ef_2}) yield:
\begin{equation}
\tilde{V}_{1,2}' \propto \sum_{m>0}\int_0^{\infty}\!dt\,t\,D_m^{12}
e^{-\Delta\hat{V}_mt}
= \sum_{m>0} \frac{D_m^{12}}{\left(\Delta\hat{V}_m\right)^2}.
\end{equation}

{}From the above formulae it is evident that not only eigenvalues of
the transfer matrix but also amplitudes enter in the computation of
the sd potentials, which illustrates where renormalization constants
come in, which are effectively removed by the multiplication with
$(g_{00}^1g_{00}^2)^{-1}$ [Eq.~(\ref{dm})].
This factor originates from the denominator of Eq.~(\ref{defcor}).
The parameters $D_m^{12}$ and $\Delta\hat{V}_m$ can be
fixed from a fit to Eq.~(\ref{eq:correl}). The hybrid potentials
$\hat{V}_m$ can in principle also be determined
independently~\cite{huntley2}. We leave this for future high precision
studies on anisotropic lattices. For the time being, we
evaluate the integrals Eqs.~(\ref{ef_1})--(\ref{ef_4}) numerically
where our interpolation method has been inspired by the
multi-exponential result of the spectral decomposition,
Eq.~(\ref{eq:correl}).

\subsection{Interpolation procedure}
In order to determine the sd potentials, one must evaluate
integrals over correlation functions [see
Eqs.~(\ref{ef_1})--(\ref{ef_4})] that depend
on the interaction time $t$ in a multi-exponential
way, Eq.~(\ref{eq:correl}). In the
following, $C_i(t)$ will denote the two point function which has to be
integrated out in order to determine $\tilde{V}_i^{(\prime)}$ at a given
value of ${\mathbf R}$. For $i=1, 2$, $C_i(t)$ will be weighted by an
additional factor $t$ [Eqs.~(\ref{ef_1})--(\ref{ef_2})]. Two
different methods of interpolating $C_i(t)$ in between the discrete
$t$-values have been adopted:

1.\ Wherever the quality of the signals allowed for reasonable
fits, the data has been fitted to a two-exponential ansatz,
\begin{equation}
C(t) = D_1\exp(-\Delta \hat{V}_1t)+D_2\exp(-\Delta \hat{V}_2t).
\end{equation}
This amounts to three parameter fits in case of $C_1$ and $C_2$
($D_1+D_2=0$) and four parameter fits for $C_3$ and $C_4$.
Unfortunately, stable fits have only been possible for the region
of small $R$ where we do not necessarily
expect the hybrid potentials to agree
with string model predictions.

2.\ Alternatively we have performed a local exponential interpolation:
\begin{equation}
C_i(t')=C_i(t)\exp\left[-B_i(t)(t'-t)\right]\quad,\quad
B_i(t)=\ln\left(\frac{C_i(t)}{C_i(t+1)}\right)
\end{equation}
for $t\leq t'<t+1$ and $C_i(t)C_i(t+1)>0$. Due to the
multi-exponential character of the correlation function
(or statistical fluctuations) the sign might change within
the given interval. Thus, for
$C_i(t)C_i(t+1)\leq 0$, we interpolate
linearly,
\begin{equation}
C_i(t')=C_i(t)+[C_i(t+1)-C_i(t)](t'-t).
\end{equation}
For $C_1(t)$ and $C_2(t)$ quadratical interpolations are performed
within the interval $0\leq t'<1$
to account for $C_1(0)=C_2(0)=0$, where we
demand continuity of the interpolating function and its derivative
at $t=1$.

The interpolation procedure is illustrated in Figs.~\ref{c1}--\ref{c4}
for some examples.
As can be seen from Fig.~\ref{c1} where $C_1(t)$ is
displayed for $R=10$, the quadratic interpolation for $t<1$ might
differ significantly from the fit. However, this region does only
weakly contribute to the potential since
$C_1(t)$ as well as $C_2(t)$ are weighted by an additional factor $t$.
In Figs.~\ref{c2a} and \ref{c2b}, $C_2(t)$ is displayed for $R=2$ and
$R=4$, respectively. $C_4(t)$ at $R=2$ changes its sign as can be seen
from Fig.~\ref{c4}.

All statistical errors have been bootstrapped. For each potential
$\tilde{V}_i^{(\prime)}$, numerical integration has been performed up to a
value $t=\tau_i$, with $\tau_i$ chosen such that the result is
stable (within statistical accuracy) under the replacement $\tau_i
\rightarrow \tau_i-1$ for all ${\mathbf R}$. Subsequently,
systematic cut-off errors have been estimated from the exponential
tail of fits to large $t$ data points but came out
to be negligible in all cases when
compared to the statistical error from the numerical integration up to
$\tau_i$. Whenever the deviations between the fit and
interpolation results turned out to be significant
(see e.g.~Fig.~\ref{c4}), we have included them as a systematic
uncertainty into the final error on the potential value.

\subsection{Renormalization and matching}
The sd and vd potentials are computed from amplitudes of
correlation functions
rather than from eigenvalues of the transfer
matrix. This gives rise to renormalizations in respect to
the corresponding continuum potentials. A different way to illustrate the
necessity of renormalization is the fact that
the color electric (magnetic) ears,
\begin{equation}
\hat{F}_{\mu\nu}=ga^2F_{\mu\nu}
=\frac{1}{2i}\left(\Pi_{\mu\nu}-\Pi_{\mu\nu}^{\dagger}\right)\left(1+
{\mathcal O}(a^2)\right),
\end{equation}
explicitly depend on the
lattice scale $a$ and, therefore, discretization.

As in the low energy regime of interest the renormalization constants
are likely to receive relevant high order corrections, we
apply the non-perturbative HM renormalization
prescription~\cite{huntley} [cf.\ Eq.~(\ref{defcor}) and
Fig.~\ref{mren}] that is geared  to ``dividing out'' the two amplitudes
$g_{00}^1$ and $g_{00}^2$ [Eqs.~(\ref{dm}), (\ref{gm})] from the naive
lattice two point function.
Terms that do not depend on a dimensionful parameter (which is
the distance between the sources, $R$, in the case of interest) will
give rise to divergencies in the continuum limit.
The idea behind the HM procedure is to systematically
remove such terms from the correlation
functions in order to arrive at residual
renormalization constants that are close to one.

In terms of perturbation theory, one can classify the related
diagrams into pure self interactions within the ears, pure interactions
within the Wilson loops, interactions between ears and the loop and
--- to higher orders --- mixes thereof. By our renormalization
procedure all divergent diagrams that consist only of ear
or Wilson loop self interactions, cancel identically. In
addition, many more complicated contributions are removed. Within
$V_2'$ -- $V_4$, all divergencies of
orders $g^6$ and less vanish while in case of $V_1'$ this holds only
true up to order\footnote{Within $\tilde{V}_1'$, gluon exchanges between the
two ears that are not cancelled by the denominator of
Eq.~(\ref{defcor}) contribute to the self energy.}
$g^4$, such that the renormalization constants
involved will only differ from identity on a three loop
[$1 +{\mathcal O}(g^6)$] or two loop
[$1+{\mathcal O}(g^4)$] level, respectively. 

The spirit of the procedure is close to the one of
Lepage-Mackenzie who suggest to construct ``tadpole improved''
operators~\cite{tadpole}.  In fact, one finds the HM renormalization
and the tadpole improvement prescriptions to render near-equal results
(within 2~\%, as compared to overall renormalization effects of 60~\%, at
our present $\beta$-values). The remaining difference between the
``tadpole improved'' operator and the HM renormalized counterpart can
be explained in terms of more complicated diagrams involving
interactions between the ears and the time-like parts of the Wilson
loop. We take the small size of this difference as an indication that
other similar higher order terms, which we have not been able to cancel
out completely, can be neglected.

Direct numerical checks of the accuracy of the approach are possible
in two ways, namely (i) by varying the lattice resolution $a$ and
testing scaling of the results\footnote{However, due to the running
of the matching constants to the full theory with the lattice scale,
residual scaling violations of about 2.7~\% for $V_1'$,
and $V_2'$ and 5.5~\% for $V_3$ and $V_4$ are
expected after having rescaled the potentials
by relative factors as large as $a^2_{2.74}/a^2_{2.96}\approx
3.2$ and $a^3_{2.74}/a^3_{2.96}\approx 5.7$, respectively.
Needless to say, that we cannot resolve such
tiny effects from our lattice data.}
and (ii) by comparing the data with the
prediction on $V_2'-V_1'$ from the Gromes
relation, Eq.~(\ref{grom}), between spin-orbit potentials and the central
potential
(which does not undergo renormalization),
\begin{equation}
V_0'(r)=V_{2,\mbox{\scriptsize ren}}'(r)-V_{1,\mbox{\scriptsize ren}}'(r) .
\end{equation}
In Fig.~\ref{v12} we check our data on $V_2'-V_1'$ in units of
the string tension $\kappa$ against the force, obtained from a fit
to the central potential,
$V_0(r)$ [Eq.~(\ref{eq_v0_3}), Table~\ref{et2}].
As can be seen, the two data sets scale beautifully onto each
other and reproduce the central force up to lattice artefacts at
small $r$.

Renormalization is not a pure lattice problem in this case.
By truncating the $1/m$ expansion of the SU(2) Lagrangian
at a given order, the ultra-violet behavior
is altered  in respect to the full theory. Therefore, the
resulting effective
Lagrangian has to be matched to full SU(2) gauge theory
at a renormalization scale
$\mu=\pi/a$ giving
rise to coefficients $c_i(\mu,m)$, connecting an SU(2)
potential at the heavy quark mass $m$, $V_i(r;m)$ to the corresponding
potential, computed in the framework of
the effective theory at scale $\mu$, e.g.\ $V_i(r;m)=c_i(\mu,m)V_i(r;\mu)$.
This problem, which becomes visible beyond the tree-level,
has been treated in a
systematic manner for sd potentials by Chen {\em et al.}~\cite{Chen}. 
The Gromes relation still remains valid [$V_0'(r)=V_2'(r;\mu)-V_1'(r;\mu)$],
and  constrains  the matching coefficients whose one loop values
have been derived in the Reference. Unlike the renormalization
constants that relate the lattice potentials to the continuum
counterparts, in this case, we can rely on perturbative results,
with the one loop correction being
an effect of just a few per cent.

\subsection{Finite volume effects}
It is well known from previous simulations of SU(2) gauge theory (cf.\
Ref.~\cite{schlicht}) that finite size effects (FSEs) on the
central potential, computed from Wilson loops, and the string tension
are negligible within the
numerical accuracy of typical lattice studies, as long as the
spatial lattice extent is kept larger than $L_{\sigma}\approx 1$~fm (and
$L_{\tau}\ge L_{\sigma}$).
However, since the critical temperature of the deconfinement phase
transition corresponds to a scale of about 0.65~fm~\cite{fingberg},
the slope of the potential decreases rapidly as the extent of the box
is reduced below about 0.8~fm. Nonetheless, one might expect the
sd potentials $V_2'$ -- $V_4$ to be affected much less by the infrared
cut-off than the central potential, as they are of much shorter range
than the latter [see Eqs.~(\ref{pertu}),
(\ref{pertu2})--(\ref{pertu4}) or Eqs.~(\ref{a111})--({\ref{a444}) of
the Appendix].

All results at $\beta=2.74$ have been obtained on a volume of $32^4$
lattice sites which is comfortably large. In addition, we have performed
simulations on a $16^4$ lattice, whose physical
extent approximately corresponds to that of a $32^4$ lattice at
$\beta=2.96$ (Table~\ref{et1}), to estimate FSEs on the
individual potentials. At $\beta=2.96$, we have determined the string
tension and parametrization of the central potential on a $48^4$
lattice which is sufficiently large to avoid serious FSEs. The
extracted scale $a=0.022$~fm
is in agreement with the expectation we
have from the data collected in Ref.~\cite{schlicht}. As expected,
we find significant FSEs on the central potential obtained on a $32^4$
lattice at
this $\beta$-value as can be seen from Fig.~\ref{v0296}. The slope is
reduced by about 35~\% on the small volume while the string tension on a
$16^4$ lattice at $\beta=2.74$ comes out to be smaller by as much as
50~\%, compared to $V=32^4$.

At $\beta=2.96$, the computer memory forced us to restrict
ourselves to the smaller volume ($V=32^4$) for the sd potentials. In the
following, we compare results obtained on the $16^4$ lattice to
$V=32^4$ results at $\beta=2.74$ to justify the assumption that the
$\beta=2.96$ sd potentials are not seriously affected by FSEs.
As can be seen from Figs.~\ref{vol1}--\ref{vol4}, for all
sd potentials FSEs are statistically
insignificant.
Differences between the constant long range contributions to
$\tilde{V}_1'$, which we do expect, are hidden within the large statistical
errors of the $\langle\langle B_i({\mathbf 0},0)E_j({\mathbf
0},t)\rangle\rangle_W$ correlators. We
conclude that though a spatial lattice extent of $0.7$~fm is too small to
extract the infinite volume central potential, it
suffices for extracting $\tilde{V}_2'$ -- $\tilde{V}_4$ as well as the short
distance contribution to $\tilde{V}_1'$.

\section{Physics results}
\label{sec4}

\subsection{Getting started: the central potential}
The lattice potential data 
$\hat{V}_0({\mathbf R})$ has been computed from smeared Wilson loops
by the method described in Ref.~\cite{schlicht}.
Our general strategy is to derive continuous parametrizations of the
lattice data points which will enable us to construct derivatives and compare
the results to theoretical expectations, such as the
exact Gromes relation [Eq.~(\ref{grom})]
or the approximate relations, Eqs.~(\ref{exv31})--(\ref{exv3}).
The continuum tree level results
on the central and sd potentials have been presented in Section~\ref{expect}
[Eqs.~(\ref{pertu0})--(\ref{pertu4})] and the corresponding lattice
expressions are derived in the Appendix
[Eqs.~(\ref{aa6}), (\ref{v2p}), (\ref{latv31})--(\ref{latv32}) and
(\ref{latdelt})], such that we can correct the
lattice data for the differences before attempting to fit them to a
continuous parametrization.

Let
\begin{equation}
\hat{V}_{\mbox{\scriptsize corr}}(R)=\hat{V}_0({\mathbf R})-
g\delta \hat{V}_0({\mathbf R}).
\label{eq_v0_2}
\end{equation}
with
\begin{equation}
\delta \hat{V}_0({\mathbf R}) =
\frac{1}{R}-4\pi G_L({\mathbf R})
\label{eq_v0_1}
\end{equation}
be the tree level corrected central potential.
The lattice gluon propagator
$G_L({\mathbf R})$ is defined in the Appendix [Eq.~(\ref{a10})].
The lattice central potential can be fitted
to the ansatz [including $g$ of Eq.~(\ref{eq_v0_2}) as a fit parameter], 
\begin{equation}
\hat{V}_{\mbox{\scriptsize corr}}(R)-\hat{V}_c
= K R -\frac{e}{R}+\frac{\hat{f}}{R^2}-\frac{\hat{d}}{R^3}
\label{eq_v0_3}
\end{equation}
with self energy $\hat{V}_c$, string tension $K$
and Coulomb coefficient $e$. The $1/R^2$ and $1/R^3$ functional
form of the corrections
corrections that account for the running of the coupling is not meant
to be physical but has just been introduced to effectively parametrize
the data within the given range of $R$-values. For the $\beta=2.74$
data we have set $\hat{d}=0$, such that we have 5 fit
parameters in this case while we allow for 6 parameters at
$\beta=2.96$. The resulting parameter values are displayed
in Table~\ref{et2}. For technical reasons (link integration),
only potential values for $R\ge \sqrt{2}$ have been obtained,
such that the fits do not include $R=1$.

In Fig.~\ref{v0}, the
potentials $\hat{V}_{\mbox{\scriptsize corr}}-\hat{V}_c$ from both
$\beta$-values are
displayed in units of the string tension, $K$, as extracted from the
fits, together with a fit curve that corresponds to the (averaged) 
values of fit parameters $e=0.257$, $f=a\hat{f}=0.0054/\sqrt{\kappa}$ and
$d=a^2\hat{d}=0.00011/\kappa$.
As can be seen, the two data
sets scale nicely onto each other which
means that the result appliesg to the continuum. Violations of
rotational invariance
are removed by our correction method, even at very small values of
$R$, and the data is well described by the parametrization over the
whole $r$ range.

\subsection{Spin dependent Potentials}
Our results on the first spin-orbit potential $V_1'$ are displayed in
Fig.~\ref{v1} (in units of the string tension). The two data sets show
approximate scaling behavior. In addition to a constant long range
contribution $-K$ as expected from Eq.~(\ref{v1ex}), we find an
attractive short range contribution that 
can be fitted to a Coulomb-like ansatz,
\begin{equation}
\tilde{V}_1'(r)=-\frac{h}{R^2}-K.
\end{equation}
For these one parameter
fits we have constrained the constant long range part to the value of
the string tension, as obtained from the central potential. We find
the values $h=0.0474(58)$ and $h=0.0439(23)$ at $\beta=2.74$ and
$\beta=2.96$ for this dimensionless parameter, respectively. Averaging
these two numbers yields $h = 0.0458(25)$.
This additional term comes out unexpected\footnote{However, from the
results of Ref.~\cite{Chen} and the Gromes relation, we can derive the
following one loop connection between results on $V_1'(r;\mu_i)$, obtained at
lattice spacings $a_1=\pi/\mu_1$ and $a_2=\pi/\mu_2$~\cite{SU3}:
$V_1'(r;\mu_2)=V_1'(r;\mu_1)-
\left(1-(\alpha(\mu_2)/\alpha(\mu_1))^{9/25}\right)V_2'(r;\mu_1)$,
which means (i) that such a contribution must exist and (ii) that its relative
weight will increase with decreasing lattice spacing. However, the
actual magnitude of this admixture is still surprising as the
prefactor of the admixture from $V_2'$ is as small as about 0.03
under a scale change by a factor two.} and
amounts to about one fifth of the $e/r$ contribution 
to the central potential.

Taking into account the running coupling improved effective
parametrization of the central potential [Eq.~(\ref{eq_v0_3})] and the
Gromes relation, we expect
\begin{equation}
\label{v2exp}
V_2'(r)=\frac{e-h}{r^2}-\frac{2f}{r^3}+\frac{3d}{r^4}
\end{equation}
as opposed to the tree level expectation Eq.~(\ref{pertu2}).
The Coulomb coupling has to be reduced by the amount that goes into
$V_1'$.

If we assume $V_1'$ to be generated exclusively from scalar-like
exchanges and neglect the possibility of pseudoscalar contributions
to $V_3$, Eq.~(\ref{exv31}) leads us to modify
Eq.~(\ref{pertu3}) and expect,
\begin{equation}
\label{v3exp}
V_3(r)=\frac{3(e-h)}{r^3}-\frac{8f}{r^4}+\frac{15d}{r^5}.
\end{equation}

In Figs.~\ref{v2} and \ref{v3}, the spin-orbit potential $V_2'$
and the spin-spin potential $V_3$ are displayed, together with
the expectations Eqs.~(\ref{v2exp}) and (\ref{v3exp}), respectively.
In both cases, we observe reasonable agreement between data and
expectations. For small values of $R$, scaling violations between
the two data sets from the
different $\beta$-values are evident as well as (in case of $V_3$)
some deviations from the expectation. As we will see in
Section~\ref{FAE}, these differences can be explained as
finite $a$ effects and understood in terms of lattice perturbation
theory.

In Fig.~\ref{v4}, the spin-spin potential $\tilde{V}_4$ is displayed in
lattice units for the two $\beta$-values.
An oscillatory behavior is observed which is similar to that of the
lattice $\delta$-function, expected on the tree
level [Eq.~(\ref{latdelt}) of the Appendix].
Moreover, the two data sets nearly coincide with each other,
in distinct violation of scaling.
Higher order corrections to the $\delta$-function which might scale
with an appropriate dimension should account for
the differences between the two data sets at small $R$.

\subsection{Finite a aspects}
\label{FAE}
In Figs.~\ref{latt_v2}--\ref{latt_v4}, we focus on the small $R$
behavior of the sd
potentials $\tilde{V}_2',\tilde{V}_3$ and $\tilde{V}_4$. We restrict
ourselves to display the $\beta=2.74$ results only which are in
qualitative agreement
with those obtained at $\beta=2.96$. In addition
to the data (with error bars),
the tree level perturbative expressions from the Appendix
are displayed [squares, Eq.~(\ref{v2p}) for $\tilde{V}_2'$,
Eqs.~(\ref{latv31}) and (\ref{latv32}) for $\tilde{V}_3$ and
Eq.~(\ref{latdelt})
for $\tilde{V}_4$). The normalization constants $c=C_F\alpha_s$ have been
obtained from fits to the first seven data points. 
All three sd potentials are
qualitatively described by these one parameter fits and
deviations of the data from a continuous curve can be
understood in terms of this lattice expectation.

The fit parameters are displayed in Table~\ref{et3}. From the analysis
of the central potential, we would expect values $c=e-h\approx 0.21$ while
the tree level lattice expectations [with lattice coupling
$\alpha_s=1/(\pi\beta)$] are
$c=0.087$ and $c=0.081$ for $\beta=2.74$ and $\beta=2.96$,
respectively. In agreement with the
perturbative expectation, all $c_i$ come out to decrease with
increasing
$\beta$. We find $c=e-h$ as determined from $V_0$ and $V_1'$ to be
about 2.5 times larger than the
naive tree level value while this factor reduces to 1.8 in case of
$\tilde{V}_2'$ and 1.3 for $\tilde{V}_3$ and $\tilde{V}_4$ as these
potentials are dominated
by higher momentum gluon exchanges and thus more perturbative.
In order to investigate if the remaining differences between data
points and tree level expectation (with renormalized coupling $c_i$
as fit parameters) can be explained in terms of higher order
perturbative corrections, we attempt to model running coupling effects.

The only additional diagrams that
contribute to $V_0$ at ${\mathcal O}(g^4)$ on the lattice as well as
in the continuum are one loop corrections to the gluon self energy.
The renormalization of the coupling, emanating from these diagrams, has
been computed on the lattice for on-axis separations of the
sources~\cite{paffuti,karsch}. One can account for this correction by
building in a running coupling constant $\alpha({\mathbf q})$ into the
gluon propagator in momentum space [Eq.~(\ref{a10})].
Instead of attempting to compute the correct lattice sum, we model
this effect by the corresponding continuum expression,
\begin{equation}\label{running}
\alpha(t)=\frac{1}{4\pi b_0t}\left(1+\frac{b}{t}\ln t
+\frac{b^2}{t^2}\ln t\right)^{-1}
\end{equation}
with
\begin{equation}
t=\ln\left(\frac{\hat{q}^2}{\Lambda^2}\right)\quad,\quad
b_0=\frac{11N}{48\pi^2}\quad,\quad
b_1=\frac{34N}{3(16\pi^2)^2}\quad,\quad
b=\frac{b_1}{b_0^2},
\end{equation}
where we replace $q^2$ by its lattice counterpart
$\hat{q}^2=4\sum_i\sin^2(q_i/2)$ to account for the periodic boundary
conditions. The case $b=0$ corresponds to the one loop result.
The difference to the correct lattice expression
of Ref.~\cite{karsch} is small. $\Lambda$ is a QCD scale
parameter that can
be related to the usual schemes via perturbation theory.

To ${\mathcal O}(g^6)$, apart from a renormalization of the gluon
propagator, additional exchange contributions appear that can be
resummed into a single running coupling by renormalization group
arguments. These arguments do not apply to the lattice where
rotational invariance is broken, such that, in addition to its absolute
value, the direction of ${\mathbf q}$ has to be taken into
account. Bearing this in mind, we will nonetheless attempt to model
two loop effects by the continuum two loop running coupling of
Eq.~(\ref{running}).

In case of the sd potentials $V_2'$--$V_4$, not only the gluon
self energy contributes to ${\mathcal O}(g^4)$ but also exchange
diagrams between the ears, incorporating a three gluon
vertex. In the continuum these can be resummed into an effective
running coupling. Due to this resummation, the scale parameters
$\Lambda_i$ (for $V_i^{(\prime)}$) can
differ from each other.
Again, we attempt to model 
this effect by plugging the continuum running coupling at scales
$\hat{q}$ into the
lattice tree level expressions.

In case of the sd potentials, the infrared region is
suppressed by powers of $q$ [Eqs.~(\ref{a222})--(\ref{a444})], such
that the form of the propagator in
the non-perturbative domain has little effect. To remove the
unphysical pole at $q=\Lambda$, an infrared
protection can be build in into the propagator by substituting $t$ by
$t_c=\ln(q^2/\Lambda^2+c^2)$ with a constant $c$.
The smallest momentum on the finite lattice is $q=\pi/(aL_{\sigma})$. We
choose $c^2=\max(0,e-\pi^2/(aL_{\sigma}\Lambda)^2)$ ($e$ is the Euler
constant) to guarantee
$t\geq 1$ with $c^2$ being negligible
at large momenta $q\approx 1/a$.

We do not attempt to fit the central potential to its perturbative
expectation because this would require us to put a $1/q^4$-like piece
into the propagator by hand to generate the string tension. Also,
as we found out, a large part of the short range structure is of
scalar nature and would have to be modeled separately. A running
coupling analysis of the central force in position space can be found in
Refs.~\cite{pot2,michael3}.

Fits of the one loop
as well as the two loop running coupling 
improved expectations (with one free parameter $\Lambda$)
to the first 4--8 data points of each potential have been
performed. The results of the two loop fits to 7 data points
are included into Figs.~\ref{latt_v2}--\ref{latt_v4} (circles, dotted
lines). The running coupling significantly improves agreement with the
data. Differences
between one and two loop results are small. The $\Lambda$
parameters remain stable against the variation of the fit range within
10~\%.

In Table~\ref{et3}, results on the $\Lambda$ parameters from our one
and two loop fits are collected. We do not include any errors since the
fits are only thought to qualitatively describe the data with reduced
$\chi^2$-values of typically 10--100. We observe approximate scaling
between the two sets of $\Lambda$ parameters at $\beta=2.74$ and
$\beta=2.96$ within 15~\%. However, the two loop values are
larger by about a factor two than the corresponding one loop values.

\section{Conclusions and outlook}
We have devised methods to determine spin dependent interquark
forces to high precision. From our high statistics lattice
simulation in SU(2) gauge theory, we find reliable renormalized
potentials with good scaling behavior.  There is clear evidence for a
short range scalar exchange contribution in the long range spin-orbit
potential at the level of 20~\% of the Coulomb part of the central
potential. The other sd potentials are found to be short ranged and
are well understood by means of perturbation theory.

An extension of the present investigations to the case of
interest, SU(3) gauge theory, and  inclusion of  velocity dependent
corrections will be presented in II~\cite{SU3}.
As a further step, predictions from various models of QCD
interactions, such as dual QCD~\cite{dqcd} can be tested against lattice
results on the potentials. Also spectra,
wave functions and decay constants for arbitrary (sufficiently large)
values of the quark masses
can be computed just by solving a simple differential equation~\cite{como}.
Subsequently, these results can be
confronted with experiment or compared to results
from direct lattice NRQCD predictions~\cite{nrqcd} as a first
principles check of the viability of the instantaneous approximation
of the potential picture.

\acknowledgements
During completion of this work GSB has been supported by EU grant ERB
CHBG CT94-0665. We appreciate support by EU grants
SC1*-CT91-0642 and CHRX-CT92-00551 as well as
DFG grants Schi 257/1-4 and Schi 257/3-2.

\appendix
\section{Weak coupling expansion of lattice potentials}
In this appendix, we will elaborate the tree level expectations for
the sd potentials from a weak coupling expansion of SU(N) gauge
theory on the lattice. Some
of these expressions have already been derived in
Ref.~\cite{huntley} for on-axis source separations.
To illustrate the method, we start with the central potential before
considering sd corrections. We include the corresponding
continuum expressions for completeness.

\subsection{Central potential}
A weak coupling expansion of the Wilson loop yields to lowest order in
$g^2$,
\begin{eqnarray}
\langle W({\mathbf R},T)\rangle&=&
\exp\left(-C_Fg^2\frac{1}{2}\sum_{m,n,\mu,\nu}J_{\mu}(m)
G(m-n)\delta_{\mu\nu}J_{\nu}(n)\right)\\\label{a123}
&=&\exp\left(C_Fg^2T\sum_{t=0}^{T-1}
\left[G({\mathbf R},t)-G({\mathbf 0},t)\right]\right)\quad (T\gg R)
\end{eqnarray}
where $J_{\mu}(m)=\pm 1$ if $(m,\mu)\in \partial W$ and
$J_{\mu}(m)=0$ elsewhere.
Only terms, extensive in $T$, have been kept in Eq.~(\ref{a123}).
For SU(2), the color factor $C_F=\frac{N^2-1}{2N}$
becomes $C_F=3/4$.
$G(n)$ denotes the lattice gluon propagator in position space,
\begin{equation}
\label{a11}
G(n)=\frac{1}{L_{\sigma}^3L_{\tau}}
\sum_{q\ne 0}\frac{e^{iqn}}{\sum_{\mu}\hat{q}_{\mu}^2}\quad,\quad
\hat{q}_{\mu}=2\sin\left(\frac{q_{\mu}}{2}\right)
\end{equation}
with
\begin{eqnarray}
q_i &=& \frac{2\pi}{L_{\sigma}} m_i\quad, \qquad
m_i=-\frac{L_{\sigma}}{2}+1,\ldots,\frac{L_{\sigma}}{2},\nonumber\\
q_4 &=& \frac{2\pi}{L_{\tau}} m_4\quad, \qquad
m_4=-\frac{L_{\tau}}{2}+1,\ldots,\frac{L_{\tau}}{2}.
\end{eqnarray}
Note, that we have neglected the zero momentum
contribution to $W$ which is suppressed by 
a factor $RT/(L_{\sigma}^3L_{\tau})$.

With
\begin{equation}
\label{a10}
G_L({\mathbf R})=\sum_{t}G({\mathbf R},t)=
\frac{1}{L_{\sigma}^3}
\sum_{{\mathbf q}\ne {\mathbf 0}}\frac{e^{i{\mathbf q}{\mathbf R}}}
{\sum_i\hat{q}_i^2}
\end{equation}
and $\hat{V}_0({\mathbf R})=
-\lim_{T\rightarrow\infty}\ln\left(W({\mathbf R},T)\right)/T$, one
obtains
\begin{equation}\label{aa6}
\hat{V}_0({\mathbf R})=-C_Fg^2\left(G_L({\mathbf R})-G_L({\mathbf
0})\right)
\longrightarrow \hat{V}_c - C_F\alpha_s\frac{1}{R}\quad (R\rightarrow\infty)
\end{equation}
for the central potential
with $\alpha_s = g^2/(4\pi)$. $\hat{V}_c=C_Fg^2G_L({\mathbf 0})$
denotes a self energy constant.

\subsection{Spin-spin potentials}
To compute $\tilde{V}'_1({\mathbf R})$
-- $\tilde{V}_4({\mathbf R})$ from
Eqs.~(\ref{ef_1}) -- (\ref{ef_4}), only gluon exchanges between the two
color field insertions (ears) have to be considered to order $g^2$.
Starting from 
\begin{equation}
\langle P_1P_2\rangle = \exp(-C_Fg^2 X_{12})
\end{equation}
with
\begin{equation}
X_{12}=\sum_{m,n,\mu,\nu}
J^1_{\mu}(m)G(m-n)\delta_{\mu\nu}J^2_{\nu}(n)
\end{equation}
where $J^i_{\mu}(n)=\pm 1$ (the sign depends on the orientation of the link)
if $(n,\mu)\in \partial P_i$ and $J^i_{\mu}(n)=0$ elsewhere,
we obtain to order $g^2$ [Eq.~(\ref{defcor})],
\begin{eqnarray}
\left\langle\left\langle
\tilde{F}_1\tilde{F}_2\right\rangle\right\rangle_W
&=&-\frac{\left\langle
{\mathcal P}\left[W\,(P_1-P^{\dagger}_1)(P_2-P^{\dagger}_2)\right]
\right\rangle\left\langle\left. W\right.\right\rangle}
{\left\langle {\mathcal
P}\left[W\,(P_1+P^{\dagger}_1)\right]\right\rangle
\left\langle {\mathcal
P}\left[W\,(P_2+P^{\dagger}_2)\right]\right\rangle}\label{a125}\\\nonumber
&=&
-\frac{1}{4}\left\langle(P_1-P^{\dagger}_1)(P_2-P^{\dagger}_2)\right
\rangle\\\nonumber
&=&
-\frac{1}{2}\left(\langle P_1P_2\rangle-\langle
P_1P_2^{\dagger}\rangle\right)\\
&=&
\sinh\left(C_Fg^2 X_{12}\right)=C_Fg^2 X_{12},
\end{eqnarray}
with $\tilde{F}_j=\frac{1}{2i}\left(P_j-P_j^{\dagger}\right)$.
Note, that the self energy contributions from interactions within one
ear are not small, though of order $g^4$, but a 60~\% contribution at our
$\beta$-values. However, such contributions are cancelled by the
denominator of Eq.~(\ref{a125}). Within the potentials $\tilde{V}_2'$ --
$\tilde{V}_4$, all order $g^6$ self interactions are cancelled as well.

Let us consider the correlation function between two magnetic ears at
positions $({\mathbf 0},0)$ and $({\mathbf R},t)$ where we choose the
first ear to be within the $\hat{i}-\hat{j}$ plane and the latter
within the $\hat{i}-\hat{k}$ plane with $i\ne j$, $i\ne k$.
In this case, we obtain
\begin{equation}\label{a124}
V_{12}=2\int_{0}^{\infty}\!dt\,\langle\langle
\tilde{F}_1\tilde{F}_2\rangle\rangle_W
=-C_Fg^2\left(\Delta_j^{(-)}\Delta_k^{(+)}
+\delta_{jk}\Delta_i^{(-)}\Delta_i^{(+)}\right)G_L({\mathbf R})
\end{equation}
where
\begin{equation}
\Delta_i^{(+)}f({\mathbf n})=f({\mathbf n}+{\mathbf \hat{i}})
-f({\mathbf n})\quad,\quad
\Delta_i^{(-)}f({\mathbf n})=f({\mathbf n})
-f({\mathbf n}-{\mathbf \hat{i}})
\end{equation}
are forward/backward differences.
Note, that
\begin{equation}
G_L({\mathbf R})=2\int_0^{\infty}\!dt\,G({\mathbf R},t).
\end{equation}

Interactions between averages of four adjacent plaquettes centered
around lattice points ${\mathbf 0}$ and ${\mathbf R}$, which we use for
the magnetic ears [Eq.~(\ref{epi})],
can be derived from Eq.~(\ref{a124}) by averaging
over the 16 possible combinations of single plaquette ears,
\begin{eqnarray}
V_{12}^A&=&
-C_Fg^2\left(\Delta_j^{(-)}\Delta_k^{(+)}
+\delta_{jk}\Delta_i^{(-)}\Delta_i^{(+)}\right)
\Xi^{(+)}_j\Xi^{(-)}_k\Xi_iG_L({\mathbf R})\\
&=&-C_Fg^2\left(\Delta_j\Delta_k\Xi_i
+\delta_{jk}\Delta_i^2\Xi_j\right)
G_L({\mathbf R})
\end{eqnarray}
with
\begin{eqnarray}
\Xi_i^{(+)} f({\mathbf n})&=&\frac{1}{2}\left(f({\mathbf n})
+f({\mathbf n}+{\mathbf \hat{i}})\right),\\
\Xi_i^{(-)} f({\mathbf n})&=&\frac{1}{2}\left(f({\mathbf n})
+f({\mathbf n}-{\mathbf \hat{i}})\right),\\
\Xi_i f({\mathbf n})&=&\frac{1}{4}\left(2f({\mathbf n})
+f({\mathbf n}+{\mathbf \hat{i}})
+f({\mathbf n}-{\mathbf \hat{i}})\right),\\
\Delta_i f({\mathbf
n})&=&\frac{1}{2}\left(f({\mathbf n}+{\mathbf \hat{i}})-
f({\mathbf n}-{\mathbf \hat{i}})\right).
\end{eqnarray}
Note, that $\Xi_i=
\frac{1}{2}(\Xi_i^{(+)}+\Xi_i^{(-)})
=\Xi_i^{(+)}\Xi_i^{(-)}$, $\Xi_i\Xi_j=\frac{1}{2}(\Xi_i+\Xi_j)$,
$\Delta_i=\frac{1}{2}(\Delta_i^{(+)}+\Delta_i^{(-)})=
\Delta^{(+)}_i\Xi^{(-)}_i=
\Delta^{(-)}_i\Xi^{(+)}_i$. Since all $\Xi$ and
$\Delta$ are linear combinations of translations, they commute with
each other.

In case of $\tilde{V}_4$ the two magnetic ears are parallel, i.e.\ $j=k$,
such that a sum over the three possible $i,j$ combinations
[Eq.~(\ref{ef_4})] yields,
\begin{equation}\label{latdelt}
\tilde{V}_4({\mathbf R})=-2C_Fg^2\sum_{i=1}^3
\Delta_i^2
\Xi_i^{(\perp)} G_L({\mathbf R})
=-2C_Fg^2\Delta^{(2)}\Xi\, G_L({\mathbf R})
\end{equation}
with
\begin{eqnarray}
\Xi^{(\perp)}_i&=&\frac{1}{2}\sum_{j\ne i}\Xi_j\quad,\quad
\Xi=\frac{1}{3}\sum_i\Xi_i,\\
\Delta_i^{(2)}&=&\Delta_i^{(+)}\Delta_i^{(-)}\quad,\quad
\Delta^{(2)}=\sum_i\Delta_i^{(2)}.
\end{eqnarray}
Note, that $\Delta_i^2=\Delta_i^{(2)}\Xi_i$.

Correlators between $B_l$ and $B_m$ are
required for computation of $\tilde{V}_3$. Since the direction of
${\mathbf B}$ is orthogonal to the
plane of the corresponding plaquettes, we find the relations, $k=l$,
$j=m$ and $i\ne j$, $i\ne k$ for $l\ne m$.
In case $(l, i, j)$ are cyclic,
$(m, i, k)$ are automatically anticyclic, such that we obtain an overall
minus sign.
We find for $j\ne k$ from Eq.~(\ref{ef_3})\footnote{Obviously, this expression
is only useful for off-axis separations where $R_j\ne 0$ and $R_k\ne
0$.},
\begin{equation}\label{latv31}
\frac{R_jR_k}{R^2}
\tilde{V}_3({\mathbf R}) =
C_Fg^2\Delta_j\Delta_k\Xi_iG_L({\mathbf R}).
\end{equation}
For $l=m=i$ we obtain:
\begin{eqnarray}
3R_{ii}\tilde{V}_3
&=&C_Fg^2\left(2\Delta_i^2\Xi_i^{(\perp)}
-\Delta_j^2(2\Xi_k-\Xi_i)
-\Delta_k^2(2\Xi_j-\Xi_i)\right)
G_L({\mathbf R})\nonumber\\\label{latv32}
&=&
C_Fg^2\frac{1}{2}\left(5\Delta_i^{(2)}\Xi-\Delta^{(2)}\Xi-\Delta_j^{(2)}\Xi_k
-\Delta_k^{(2)}\Xi_j\right)
G_L({\mathbf R})
\end{eqnarray}
with $j\ne i$, $k\ne i$, $j\ne k$, $R_{ij}=R_iR_j/R^2-\delta_{ij}/3$.

It is easy to see that the above expressions amount to
\begin{equation}
V_4(r)=-2C_F\alpha_s\nabla^2\frac{1}{r}=8\pi C_F\alpha_s \delta^3(r)
\end{equation}
and
\begin{equation}
V_3(r)=C_F\alpha_s\frac{r^2}{r_ir_j}\partial_i\partial_j\frac{1}{r}=
3C_F\alpha_s\frac{1}{r^3}
\end{equation}
or
\begin{equation}
V_3(r)=C_F(R_{ii})^{-1}\frac{1}{3}
\left(2\partial_i^2-\partial_j^2-\partial_k^2\right)\frac{1}{r}
=C_F\alpha_s(R_{ii})^{-1}\frac{3r_i^2-r^2}{r^5}
=3C_F\alpha_s\frac{1}{r^3}
\end{equation}
in the continuum limit.

\subsection{Spin-orbit potentials}
For computation of $\tilde{V}_1'$ and $\tilde{V}_2'$ one has to take into
account correlators between plaquettes in the
$\hat{i}-\hat{j}$ plane
and the $\hat{i}-\hat{4}$ plane. To lowest order in $g$ only exchanges
between the links oriented in $\hat{i}$ direction have to be taken
into account. With
\begin{equation}
2\int_0^{\infty}\!dt\,t\,[G({\mathbf R},t)-G({\mathbf R},t+1)]
=G_L({\mathbf R}),
\end{equation}
we obtain
\begin{equation}
W_{12}=\int_{0}^{\infty}\!dt\,t\,\langle\langle
\tilde{F}_1\tilde{F}_2\rangle\rangle_W
=-C_Fg^2\frac{1}{2}\Delta_i^{(-)}G_L({\mathbf R})
\end{equation}
for the integrated correlation function.
Averaging over the relevant plaquette combinations finally yields,
\begin{equation}
W_{12}^A=-C_Fg^2\frac{1}{2}\Delta_j\Xi_iG_L({\mathbf R}).
\end{equation}
Thus, $\tilde{V}_1'$ vanishes to order $g^2$ while the leading order
expression for the second spin-orbit potential [Eq.~(\ref{ef_2})] is,
\begin{equation}
\label{v2p}
\tilde{V}_2'({\mathbf R})=-\frac{R}{R_j}C_Fg^2\Delta_j\Xi_j^{(\perp)}
G_L({\mathbf R}).
\end{equation}

In the continuum this amounts to,
\begin{equation}
V_2'(r)=-C_F\alpha_s\frac{r}{r_j}\partial_j\frac{1}{r}=C_F\alpha_s\frac{1}{r^2}.
\end{equation}

\subsection{Tree level relations}
Tree level relations between the lattice potentials that are analogous
to Eqs.~(\ref{grom}) and (\ref{exv3}) can be derived:
\begin{eqnarray}
\label{lgro}
\tilde{V}_2'({\mathbf R})
&=&\frac{R}{R_j}\Delta_j\Xi_j^{(\perp)} \hat{V}_0({\mathbf R})
+\tilde{V}_1'({\mathbf R}),\\
\tilde{V}_4({\mathbf R})&=&2\Delta_i^{(2)}\Xi\,\hat{V}_0({\mathbf
R})=2\sum_i\Delta_i\left(\frac{R_i}{R}\tilde{V}_2'({\mathbf R})\right).
\end{eqnarray}

From Eq.~(\ref{lgro}) one might attempt to generalize
the Gromes relation.
Let us assume for the moment that a
linear difference operator exists, such that
\begin{equation}
\tilde{V}_2'({\mathbf R})
-\tilde{V}_1'({\mathbf R})
=\sum_{\mathbf n}c({\mathbf n})\hat{V}_0({\mathbf R}+{\mathbf n})
\end{equation}
with constants $c({\mathbf n})$. Both sides of the above equation can
be expanded in orders of $g^2$. At order $g^2$ we find,
\begin{equation}
\sum_{\mathbf n}c({\mathbf n})G_L({\mathbf R}+{\mathbf n})
=\frac{R}{R_j}\Delta_j\Xi_j^{(\perp)}G_L({\mathbf R}).
\end{equation}
The factor $R/R_j$ illustrates that $c({\mathbf n})$
has to depend on ${\mathbf R}$, in contradiction to the
ansatz, i.e.\ nonlinear corrections have to be included. Also,
in our numerical studies we
find the tree level relation Eq.~(\ref{lgro})
to be substantially violated at small $R$. Of course, the continuum
Gromes relation as well as the above lattice version are
retrieved at large $R$.

\subsection{Continuum results}
For continuum potentials one
obtains the following tree level expressions,
\begin{eqnarray}\label{a111}
V_0(r)&=&-C_F\alpha_s\int\frac{dq^3}{2\pi^2}\frac{e^{i{\mathbf
qr}}}{q^2}
=-C_F\frac{2\alpha_s}{\pi}\int_0^{\infty}\!dq\,\frac{\sin qr}{qr}
=-C_F\frac{\alpha_s}{r},\\\label{a222}
V_2'(r)&=&-iC_F\alpha_s\int\frac{dq^3}{2\pi^2}\frac{{\mathbf qr}}{q^2r}
e^{i{\mathbf qr}}
=-C_F\frac{2\alpha_s}{\pi}\int_0^{\infty}\!dq\,q^2r\, j_1(qr)
=C_F\frac{\alpha_s}{r^2},
\\\label{a333}
V_3(r)&=&-C_F\alpha_s\int\frac{dq^3}{2\pi^2} \frac{({\mathbf
qr})^2}{q^2r^2}e^{i{\mathbf qr}}
=-C_F\frac{2\alpha_s}{\pi}\int_0^{\infty}\!dq\,q^2j_2(qr)
=3C_F\frac{\alpha_s}{r^3},\\\label{a444}
V_4(r)&=&C_F\alpha_s\int\frac{dq^3}{2\pi^2} e^{i{\mathbf qr}}
=C_F\frac{2\alpha_s}{\pi}\int_0^{\infty}\!dq\,q^2\frac{\sin qr}{qr}
=8\pi C_F\alpha_s\delta^3(r).
\end{eqnarray}
A linear confining contribution can be
introduced by adding a $1/q^4$ term to
$V_0$ in momentum space.
The integrals for the sd potentials are
suppressed at low $q$ like $q^2$ or $q^3$,
such that we naively expect perturbation
theory to be more reliable in this case than for the ground state
potential. Also, finite size effects are expected to be smaller.
$V_1'$ vanishes at tree level.

\begin{table}
\caption{Simulation parameters. The physical scale has been obtained
from $\sqrt{\kappa}=440$~MeV.}
\label{et1}
\begin{tabular}{ccccc}
$\beta$&$V=L_{\sigma}^3L_{\tau}$&$a/\mbox{fm}$&$L_{\sigma}/\mbox{fm}$&
$n_{\mbox{\scriptsize conf}}$\\\hline
2.74&$16^4$&0.041&0.65&1290\\
2.74&$32^4$&0.041&1.30&200\\
2.96&$32^4$&0.022&0.70&204\\
2.96&$48^4$&0.022&1.05&67\\
\end{tabular}
\end{table}

\begin{table}
\caption{Fit parameters to the central potential
[Eqs.~(\ref{eq_v0_2})--(\ref{eq_v0_3})].}
\label{et2}
\begin{tabular}{cccc}
Parameter&$\beta=2.74$&$\beta=2.96$&Average value\\\hline
$K$  &0.00785(13)&0.00246 (8)& ---\\
$e$  &0.2611(35) &0.2507(41) & 0.2567(74)\\
$\hat{f}$&0.0598(33) &0.1114(76) & $0.00541(38)/\sqrt{K}$\\
$\hat{d}$& --- &0.0437(50)&$0.000108(12)/K$\\
$\hat{V}_c$&0.4898(12) &0.4334 (9) & ---\\
$g$  &0.1707(12)&0.1409 (9)& ---\\
\end{tabular}
\end{table}

\begin{table}
\caption{The tree level constants $c=C_F\alpha_s$ and $\Lambda$
parameters from the running coupling analysis of
$V_2'$, $V_3$ and $V_4$.}
\label{et3}
\begin{tabular}{cccc}
tree level&$c_2$&$c_3$&$c_4$\\\hline
$\beta=2.74$&0.17&0.12&0.13\\
$\beta=2.96$&0.14&0.10&0.11\\\hline
running $\alpha$&$\Lambda_2/\sqrt{\kappa}$&$\Lambda_3/\sqrt{\kappa}$&
$\Lambda_4/\sqrt{\kappa}$\\\hline
1-loop $\beta=2.74$&0.29&0.12&0.08\\
1-loop $\beta=2.96$&0.25&0.10&0.07\\\hline
2-loop $\beta=2.74$&0.67&0.32&0.20\\
2-loop $\beta=2.96$&0.61&0.29&0.18\\
\end{tabular}
\end{table}

\begin{figure}
\unitlength 1cm
\begin{center}
\begin{picture}(12,6)
\put(0,0){\epsfxsize=12cm\epsfbox{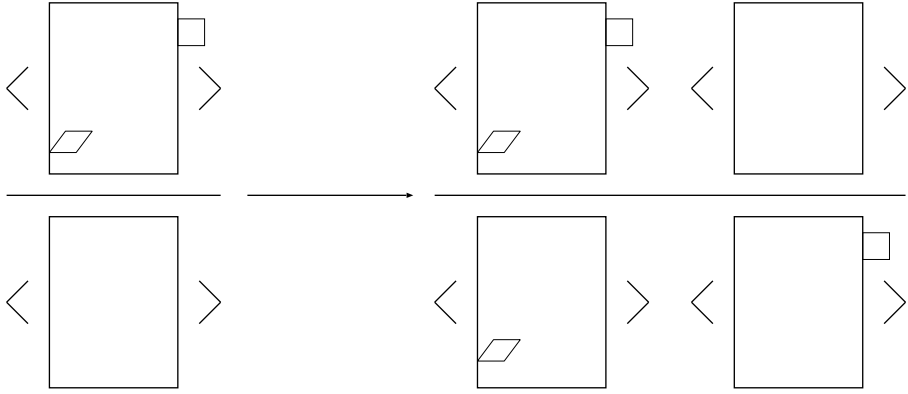}}
\end{picture}
\end{center}
\caption{Nonperturbative renormalization prescription for eared
Wilson loops.}
\label{mren}
\end{figure}

\begin{figure}
\unitlength 1cm
\begin{center}
\begin{picture}(11,8)
\put(0,0){\epsfysize=8cm\epsfbox{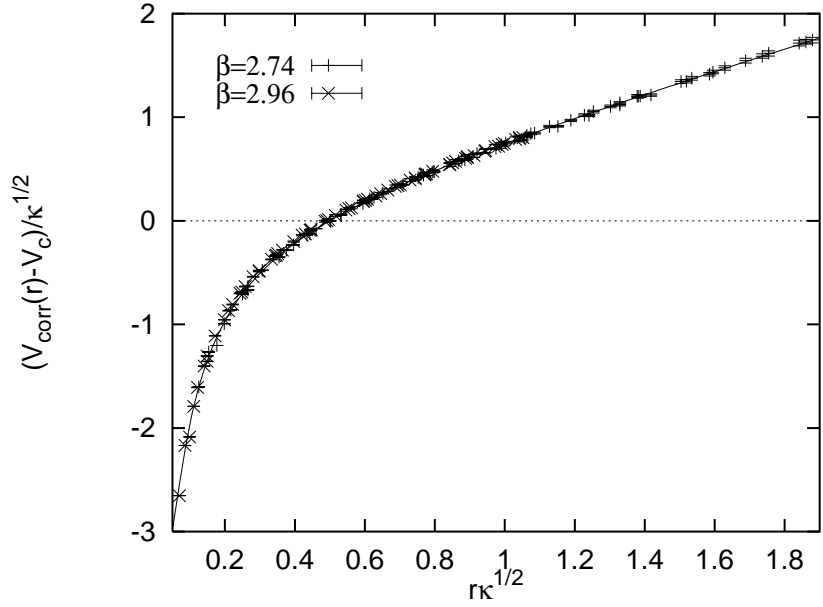}}
\end{picture}
\end{center}
\caption{Corrected central potential $V_{\mbox{\scriptsize corr}}-V_c$
in units of the string tension at $\beta=2.74$ and $\beta=2.96$.
The fit curve corresponds to the parametrization Eq.~(\ref{eq_v0_3})
with parameter values as in the last column of Table~\ref{et2}.}
\label{v0}
\end{figure}

\begin{figure}
\unitlength 1cm
\begin{center}
\begin{picture}(11,8)
\put(0,0){\epsfysize=8cm\epsfbox{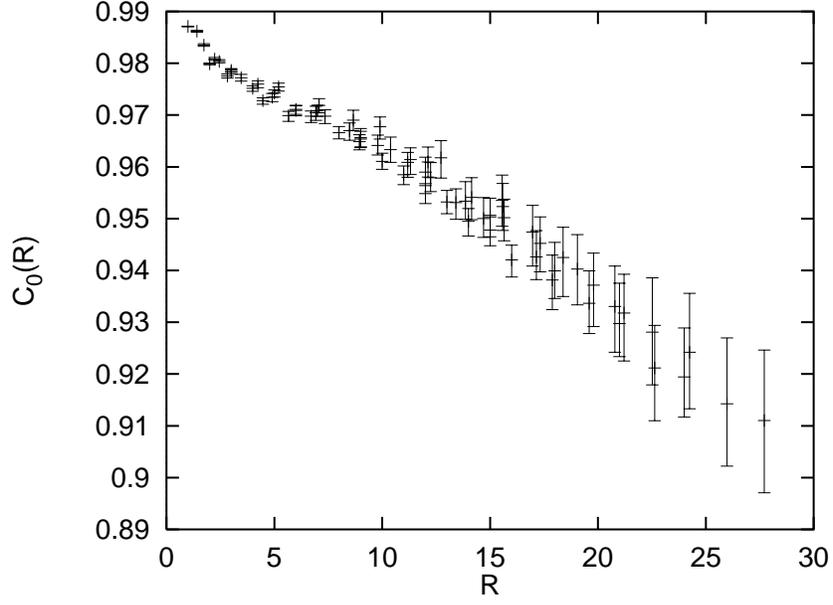}}
\end{picture}
\end{center}
\caption{Ground state overlaps as a function of the quark
separation $R$ (in lattice units) at $\beta=2.74$.}
\label{overlaps}
\end{figure}

\begin{figure}
\unitlength 1cm
\begin{center}
\begin{picture}(11,8)
\put(0,0){\epsfxsize=11cm\epsfbox{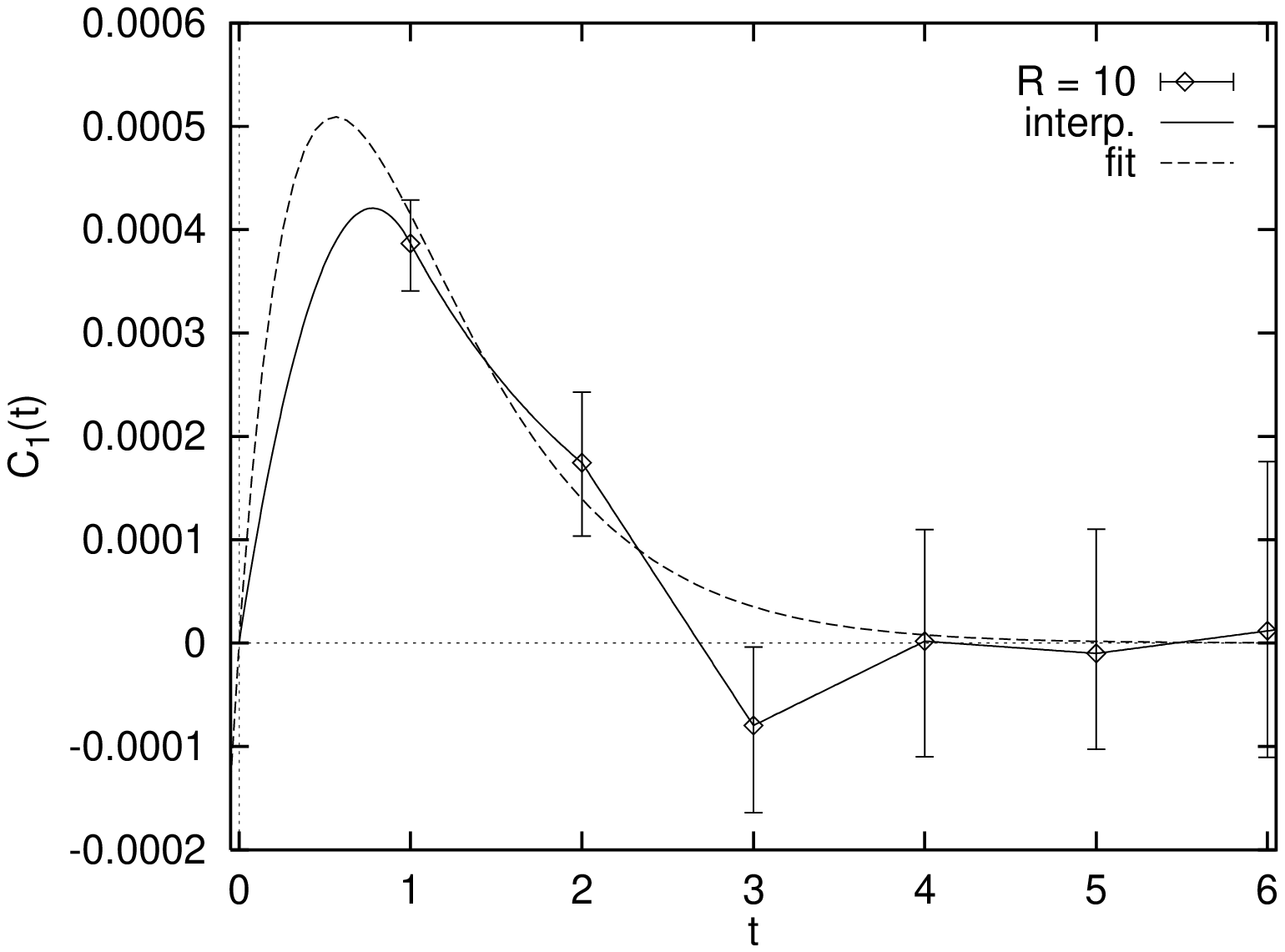}}
\end{picture}
\end{center}
\caption{The correlation function $C_1(t)$ for $R=10$. The solid curve
denotes our interpolation while the dashed curve corresponds to a
three parameter fit to the data.}
\label{c1}
\end{figure}

\begin{figure}
\unitlength 1cm
\begin{center}
\begin{picture}(11,8)
\put(0,0){\epsfysize=8cm\epsfbox{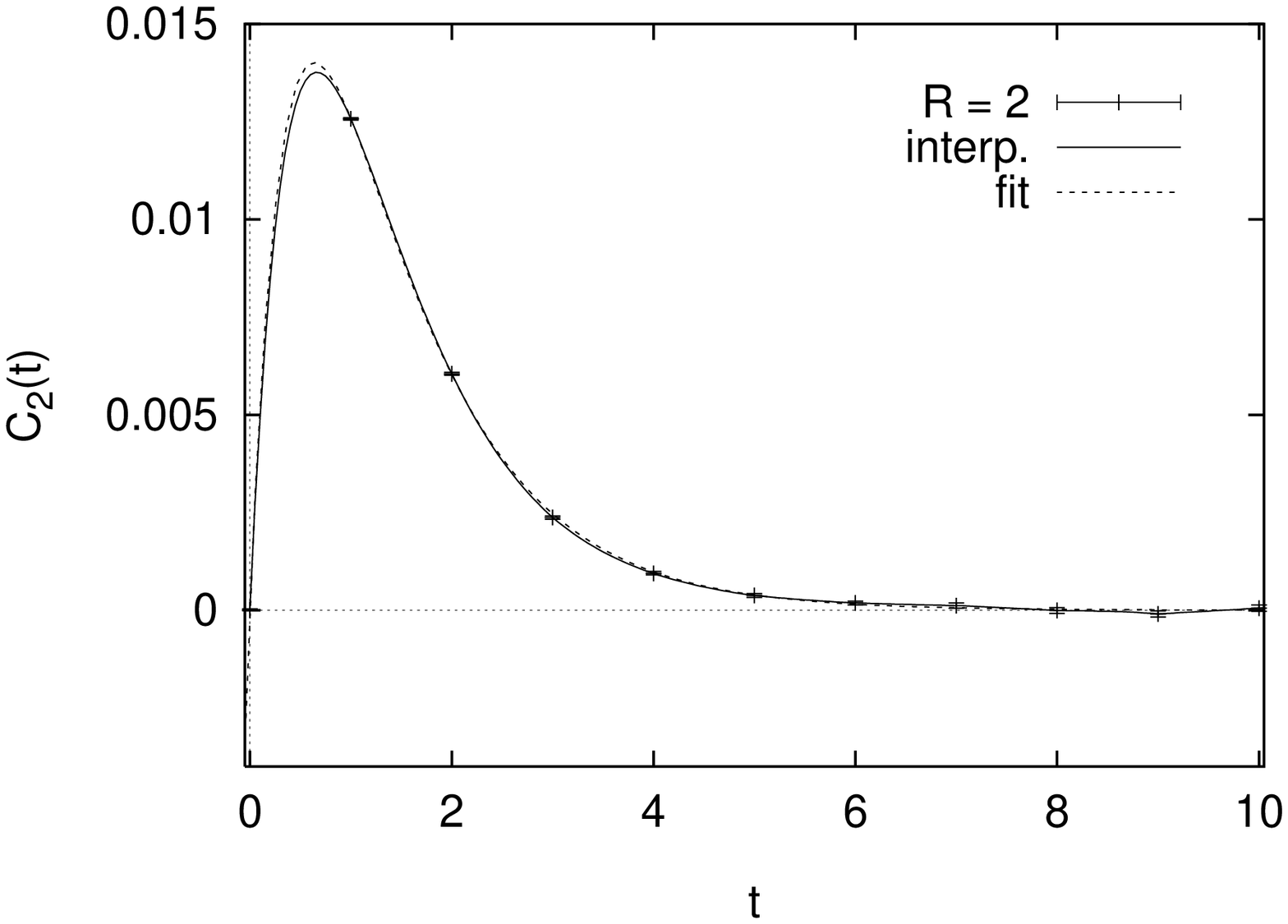}}
\end{picture}
\end{center}
\caption{The correlation function $C_2(t)$ for $R=2$. The solid curve
denotes our interpolation while the dashed curve corresponds to a
three parameter fit to the data.}
\label{c2a}
\end{figure}

\begin{figure}
\unitlength 1cm
\begin{center}
\begin{picture}(11,8)
\put(0,0){\epsfysize=8cm\epsfbox{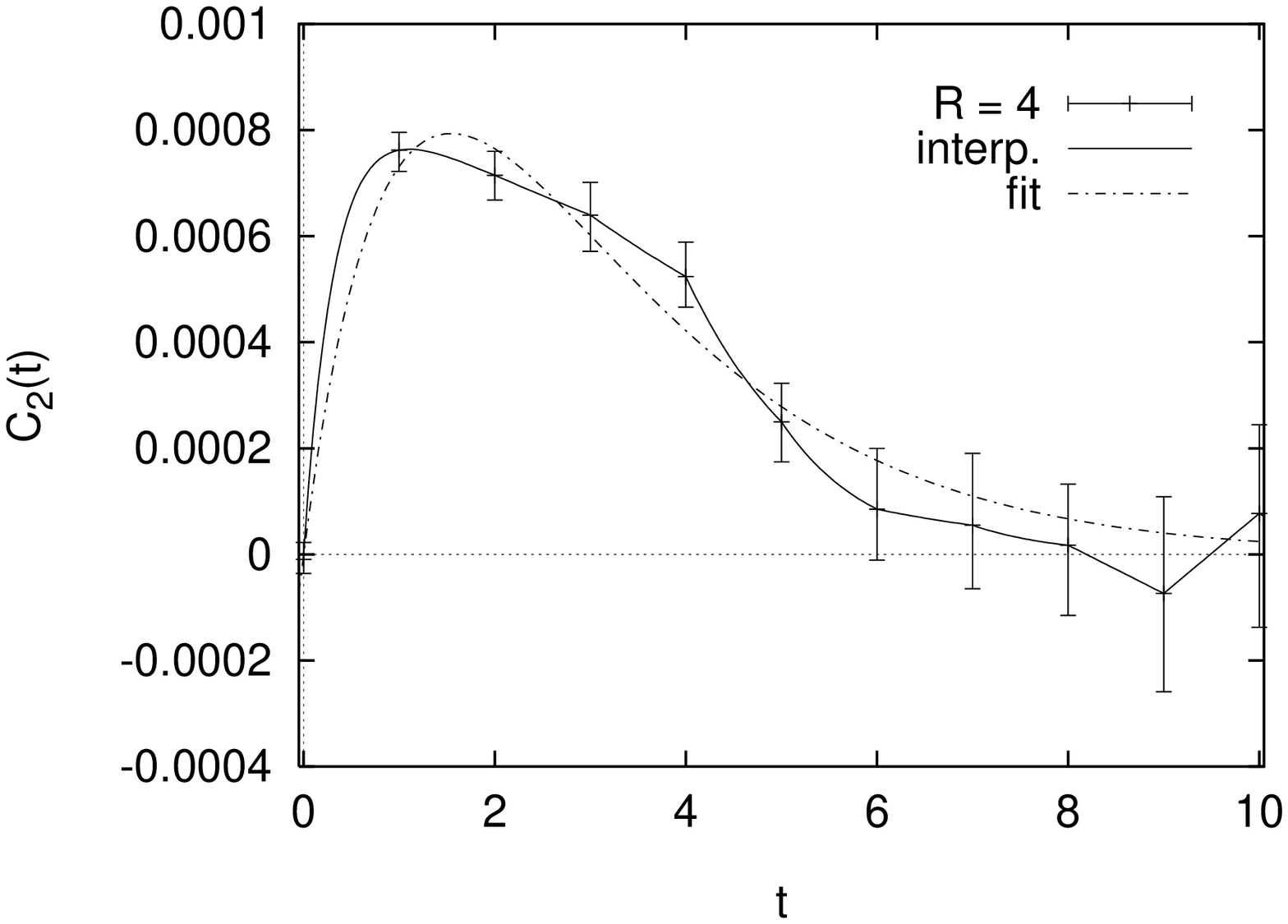}}
\end{picture}
\end{center}
\caption{The correlation function $C_2(t)$ for $R=4$. The solid curve
denotes our interpolation while the dashed-dotted curve corresponds to a
three parameter fit to the data.}
\label{c2b}
\end{figure}

\begin{figure}
\unitlength 1cm
\begin{center}
\begin{picture}(11,8)
\put(0,0){\epsfysize=8cm\epsfbox{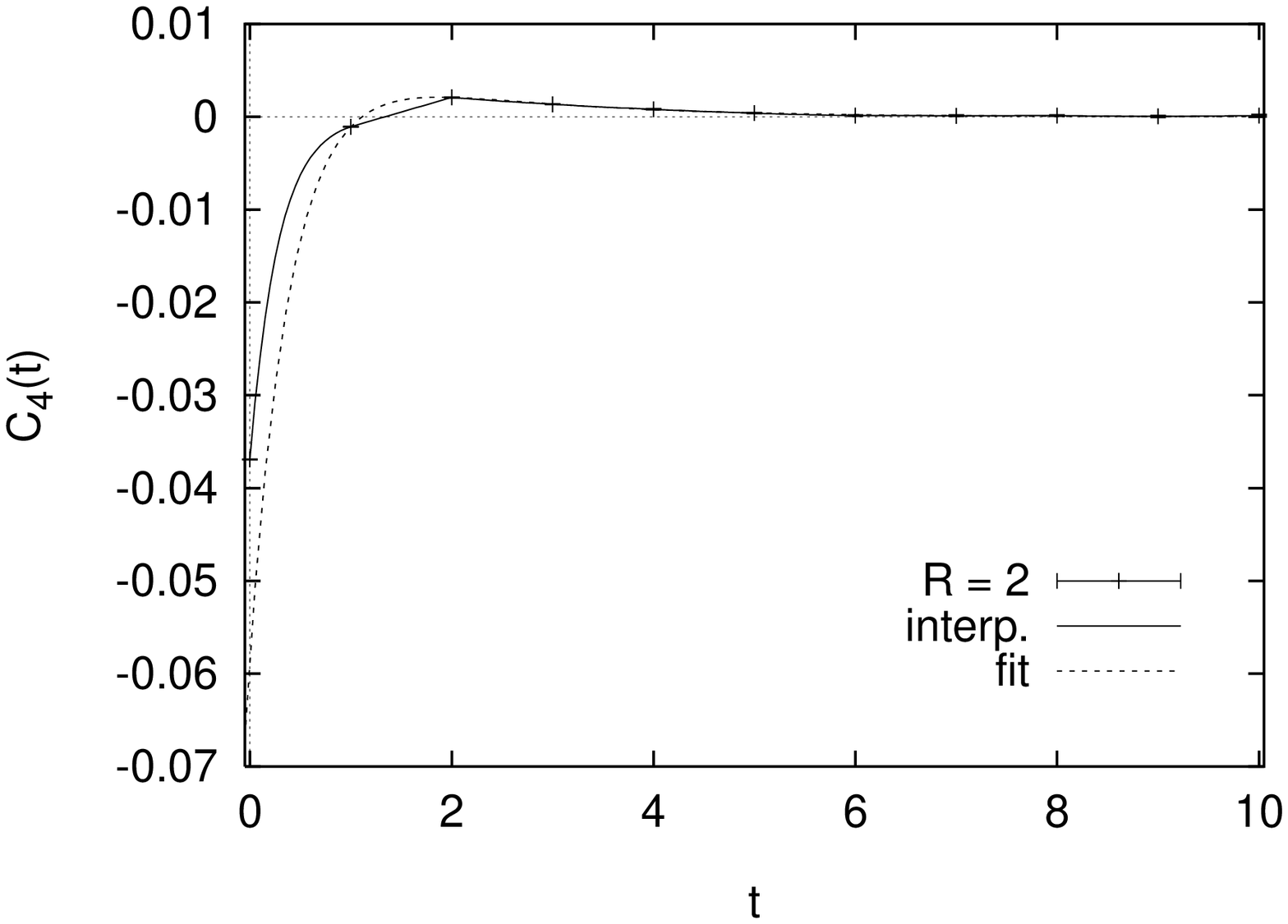}}
\end{picture}
\end{center}
\caption{The correlation function $C_4(t)$ for $R=2$. The solid curve
denotes our interpolation while the dashed curve corresponds to a
four parameter fit to the data.}
\label{c4}
\end{figure}

\begin{figure}
\unitlength 1cm
\begin{center}
\begin{picture}(11,8)
\put(0,0){\epsfysize=8cm\epsfbox{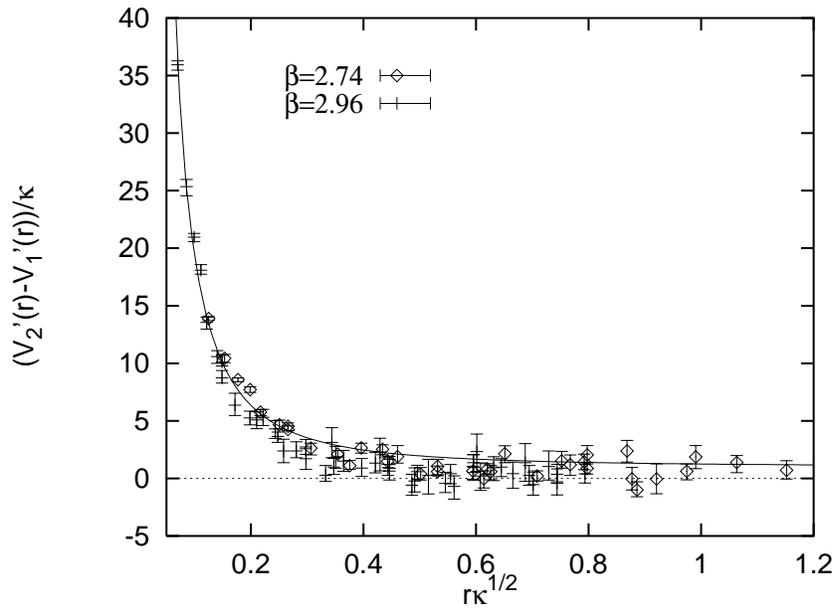}}
\end{picture}
\end{center}
\caption{Test of the Gromes relation Eq.~(\ref{grom}). The combination
$V_2'-V_1'$ is compared to the central force as obtained from the
parametrization Eq.~(\ref{eq_v0_3}).}
\label{v12}
\end{figure}

\begin{figure}
\unitlength 1cm
\begin{center}
\begin{picture}(11,8)
\put(0,0){\epsfysize=8cm\epsfbox{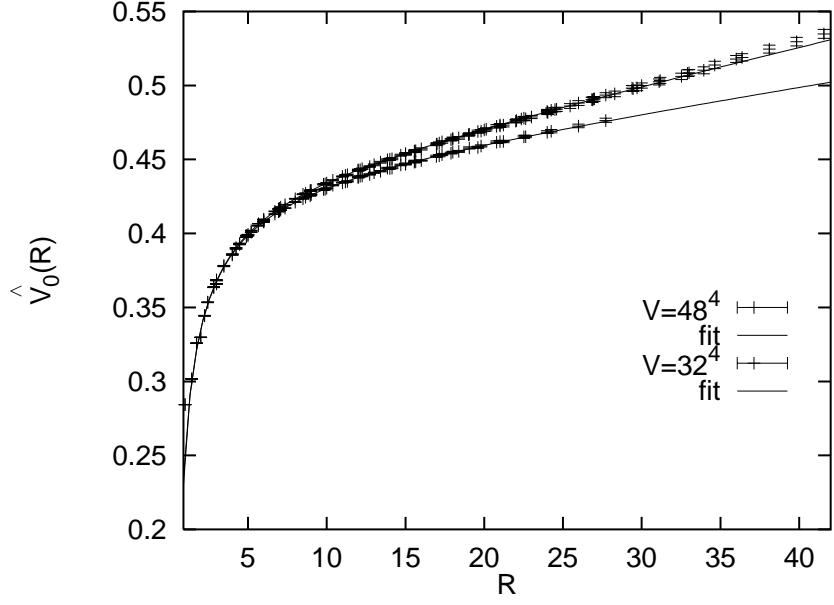}}
\put(-0.1,4.05){\mbox{\tiny$<$}}
\end{picture}
\end{center}
\caption{Comparison of the central potential $\hat{V}_0(R)$ at
$\beta=2.96$ between a $32^4$ and a $48^4$ lattice (in lattice units).}
\label{v0296}
\end{figure}

\begin{figure}
\unitlength 1cm
\begin{center}
\begin{picture}(11,8)
\put(0,0){\epsfysize=8cm\epsfbox{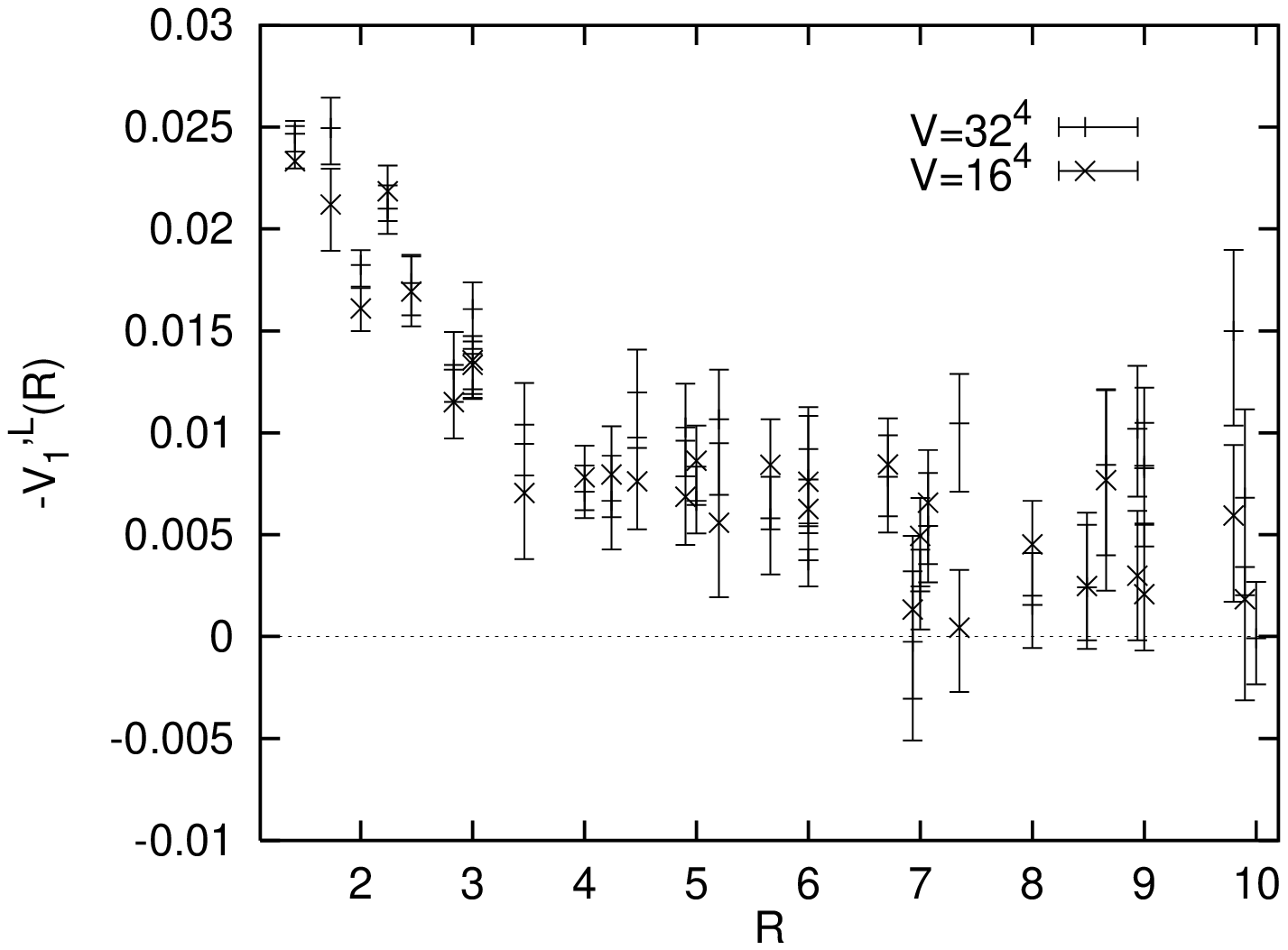}}
\end{picture}
\end{center}
\caption{Comparison between $-\tilde{V}_1'(R)$ at $\beta=2.74$ obtained on
a $16^4$ lattice and a $32^4$ lattice (in lattice units).}
\label{vol1}
\end{figure}

\begin{figure}
\unitlength 1cm
\begin{center}
\begin{picture}(11,8)
\put(0,0){\epsfysize=8cm\epsfbox{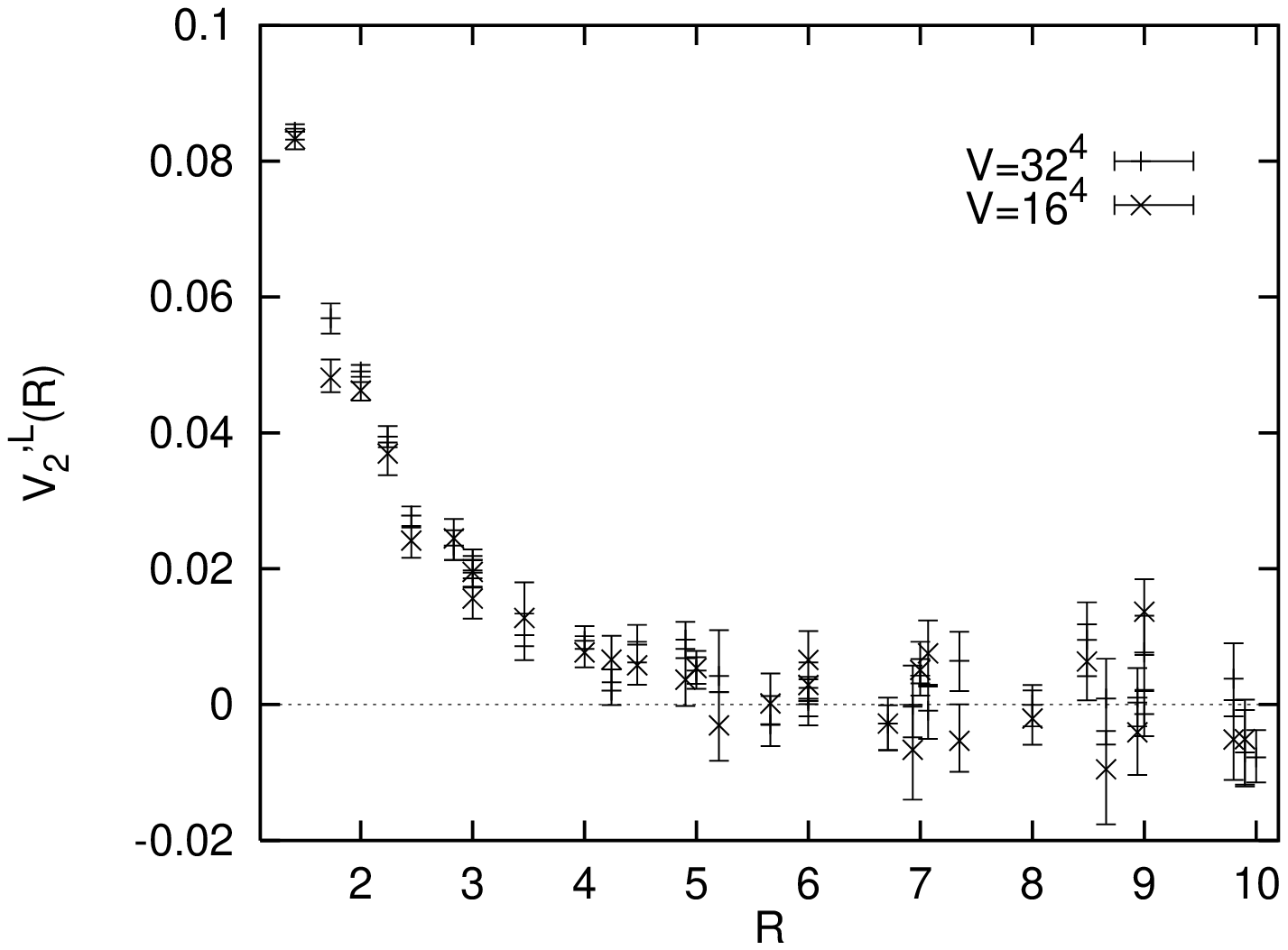}}
\end{picture}
\end{center}
\caption{Comparison between $\tilde{V}_2'(R)$ at $\beta=2.74$ obtained on
a $16^4$ lattice and a $32^4$ lattice (in lattice units).}
\label{vol2}
\end{figure}

\begin{figure}
\unitlength 1cm
\begin{center}
\begin{picture}(11,8)
\put(0,0){\epsfysize=8cm\epsfbox{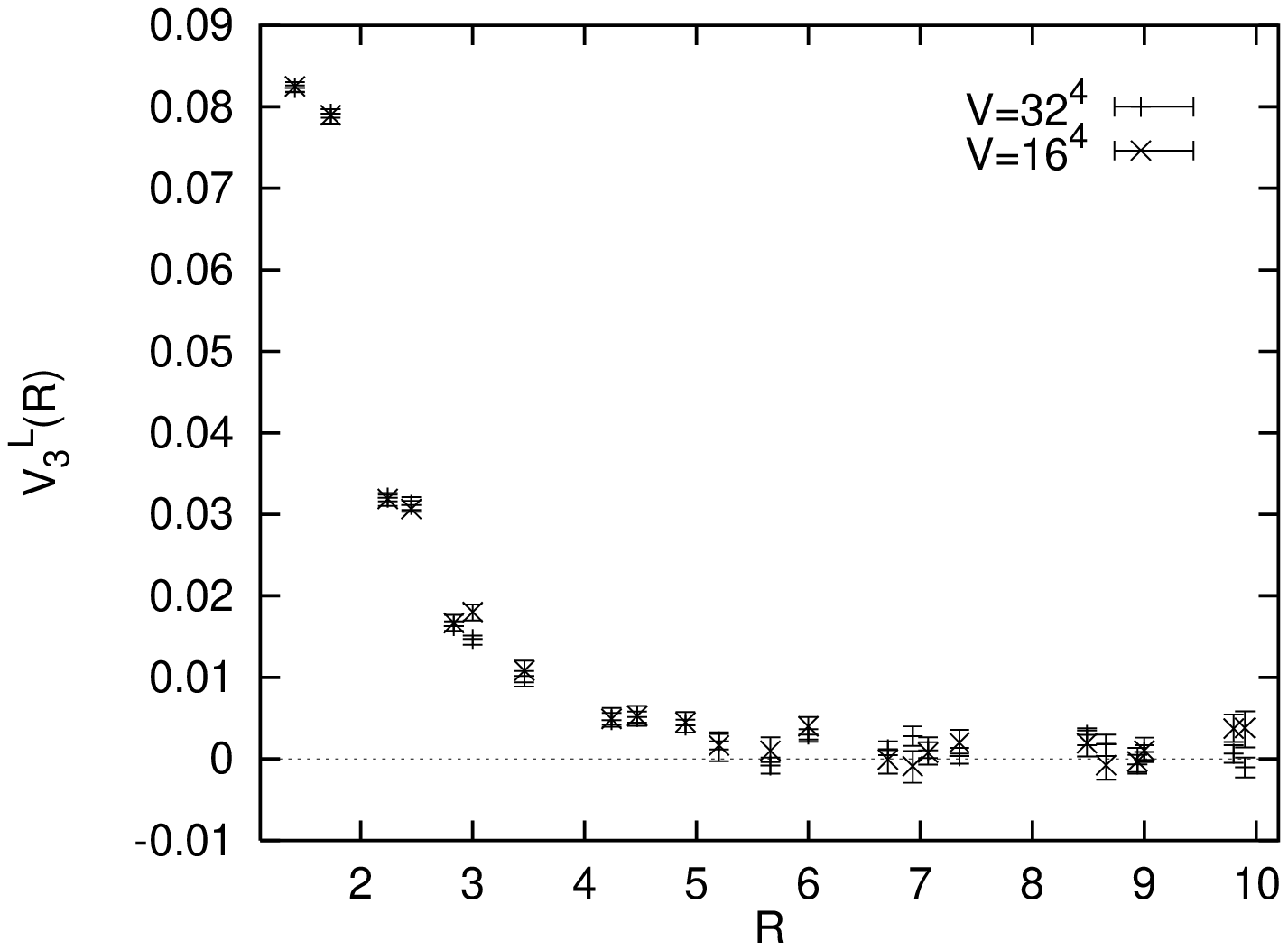}}
\end{picture}
\end{center}
\caption{Comparison between $\tilde{V}_3(R)$ at $\beta=2.74$ obtained on
a $16^4$ lattice and a $32^4$ lattice (in lattice units).}
\label{vol3}
\end{figure}

\begin{figure}
\unitlength 1cm
\begin{center}
\begin{picture}(11,8)
\put(0,0){\epsfysize=8cm\epsfbox{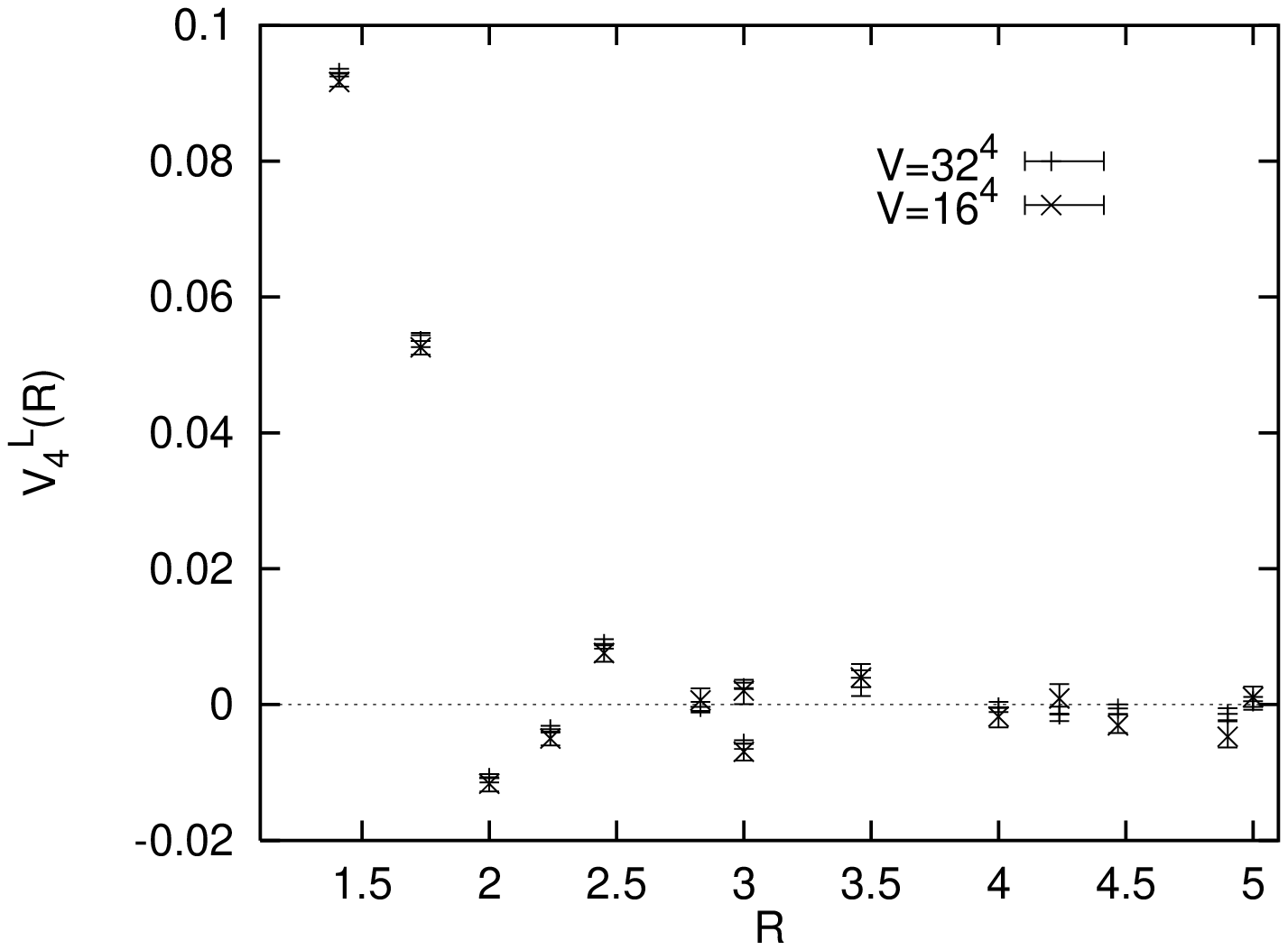}}
\end{picture}
\end{center}
\caption{Comparison between $\tilde{V}_4(R)$ at $\beta=2.74$ obtained on
a $16^4$ lattice and a $32^4$ lattice (in lattice units).}
\label{vol4}
\end{figure}

\begin{figure}
\unitlength 1cm
\begin{center}
\begin{picture}(11,8)
\put(0,0){\epsfysize=8cm\epsfbox{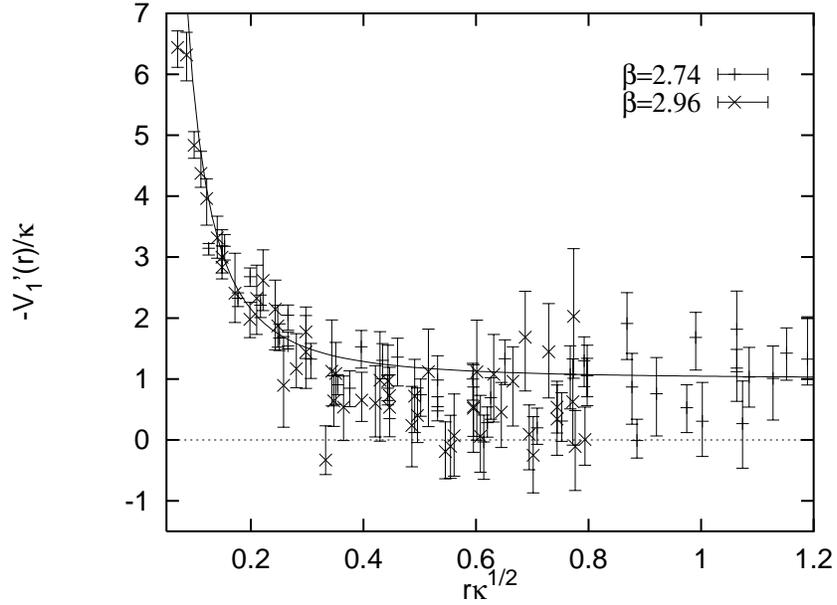}}
\end{picture}
\end{center}
\caption{The spin-orbit potential $V_1'$, together with
a fit curve of the form $-V_1'(r)=\kappa+h/r^2$ (with $h=0.046$)
in units of the string tension.}
\label{v1}
\end{figure}

\begin{figure}
\unitlength 1cm
\begin{center}
\begin{picture}(11,8)
\put(0,0){\epsfysize=8cm\epsfbox{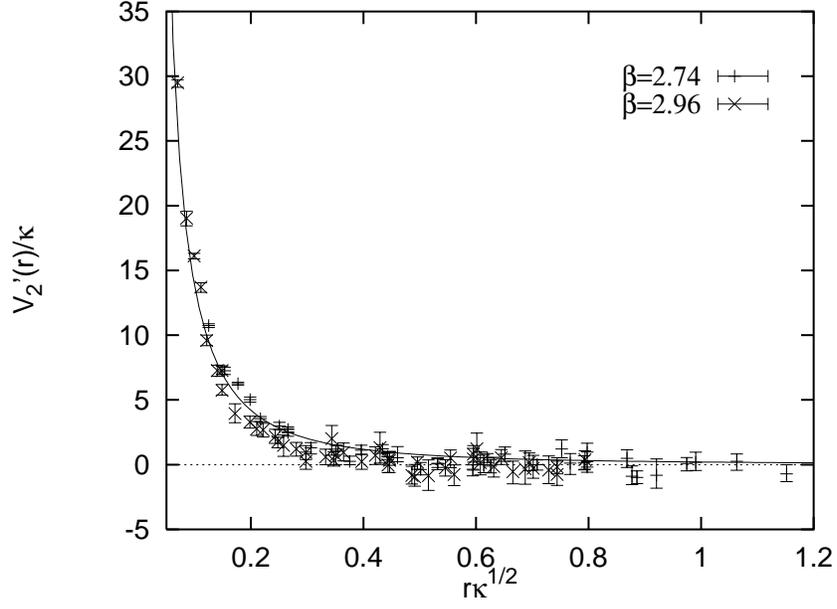}}
\end{picture}
\end{center}
\caption{The spin-orbit potential
$V_2'$ in comparison to the continuum expectation from
Eq.~(\ref{v2exp}).}
\label{v2}
\end{figure}

\begin{figure}
\unitlength 1cm
\begin{center}
\begin{picture}(11,8)
\put(0,0){\epsfysize=8cm\epsfbox{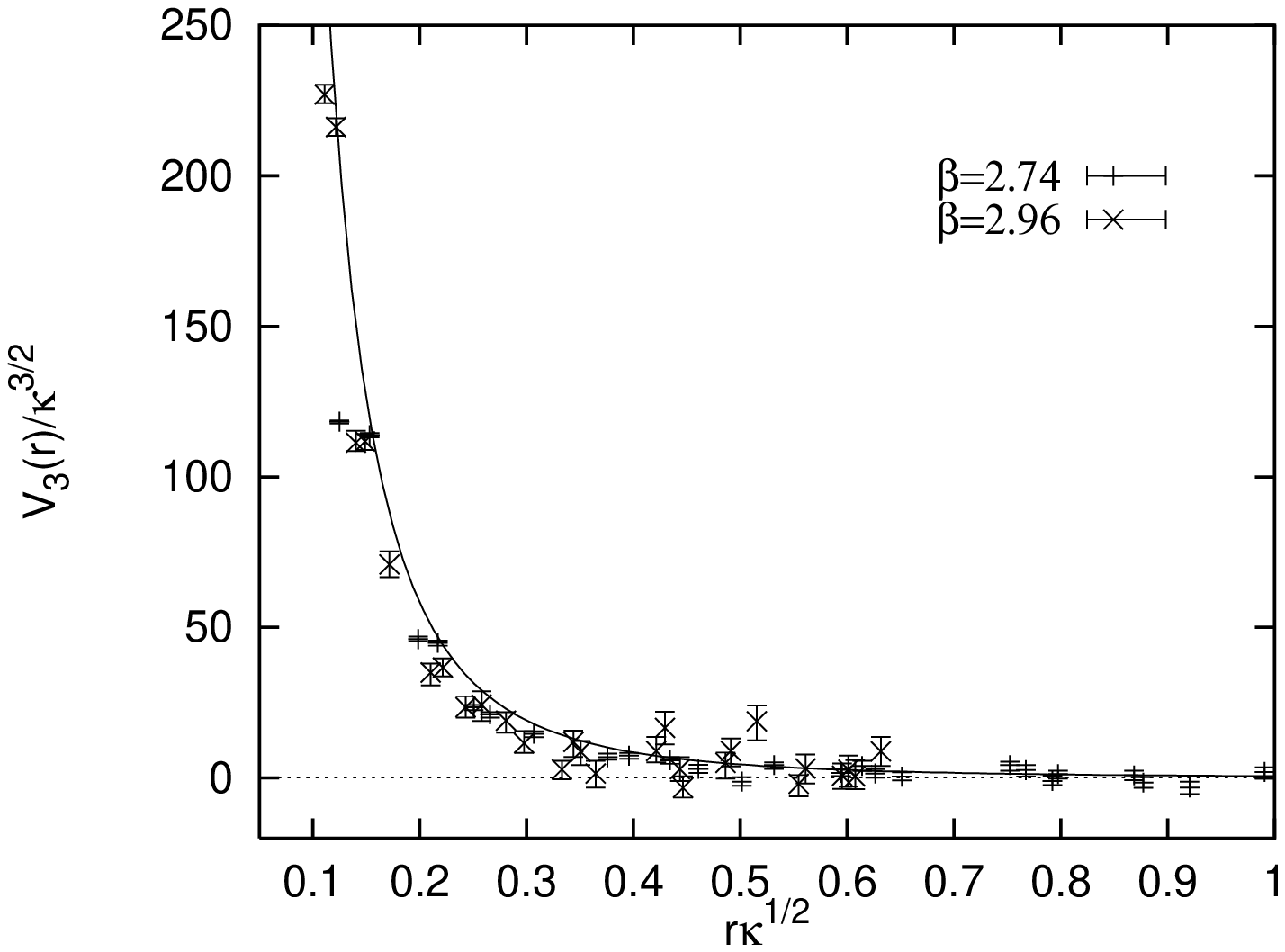}}
\end{picture}
\end{center}
\caption{The spin-spin potential
$V_3$ in comparison to the continuum expectation from
Eq.~(\ref{v3exp}).}
\label{v3}
\end{figure}

\begin{figure}
\unitlength 1cm
\begin{center}
\begin{picture}(11,8)
\put(0,0){\epsfysize=8cm\epsfbox{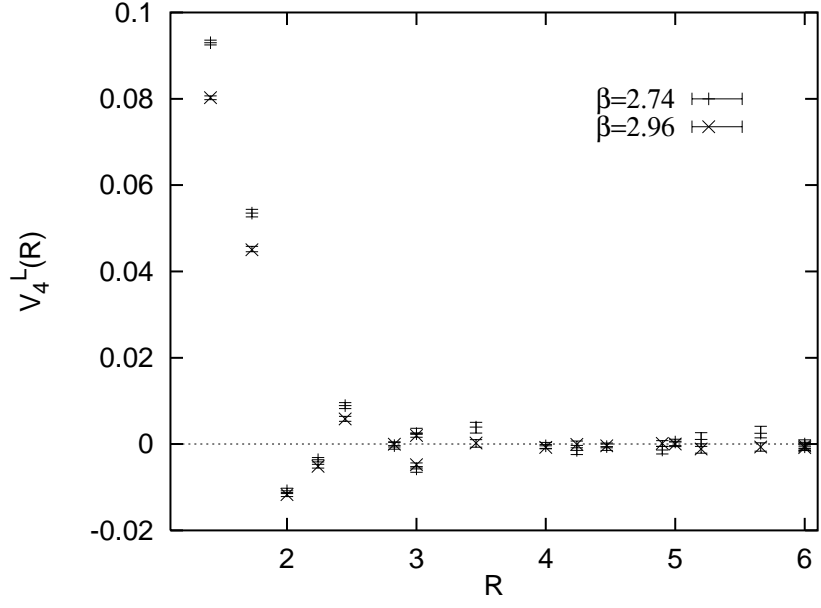}}
\end{picture}
\end{center}
\caption{The spin-spin potential $\tilde{V}_4$ for the two $\beta$-values in
lattice units.}
\label{v4}
\end{figure}

\begin{figure}
\unitlength 1cm
\begin{center}
\begin{picture}(11,8)
\put(0,0){\epsfysize=8cm\epsfbox{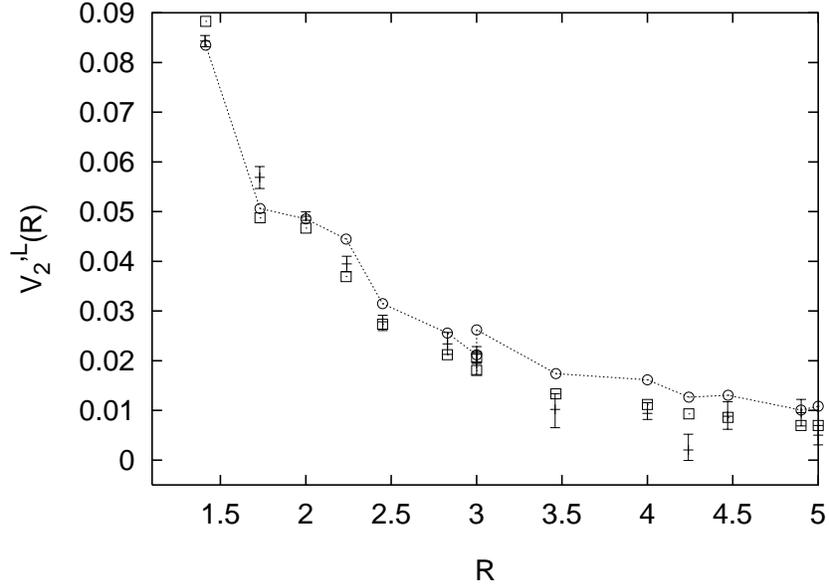}}
\end{picture}
\end{center}
\caption{Comparison of the lattice potential $\tilde{V}_2'$ (points
  with error bars) to tree level (open squares) and two loop running
  coupling improved (open circles) lattice perturbation theory.}
\label{latt_v2}
\end{figure}

\begin{figure}
\unitlength 1cm
\begin{center}
\begin{picture}(11,8)
\put(0,0){\epsfysize=8cm\epsfbox{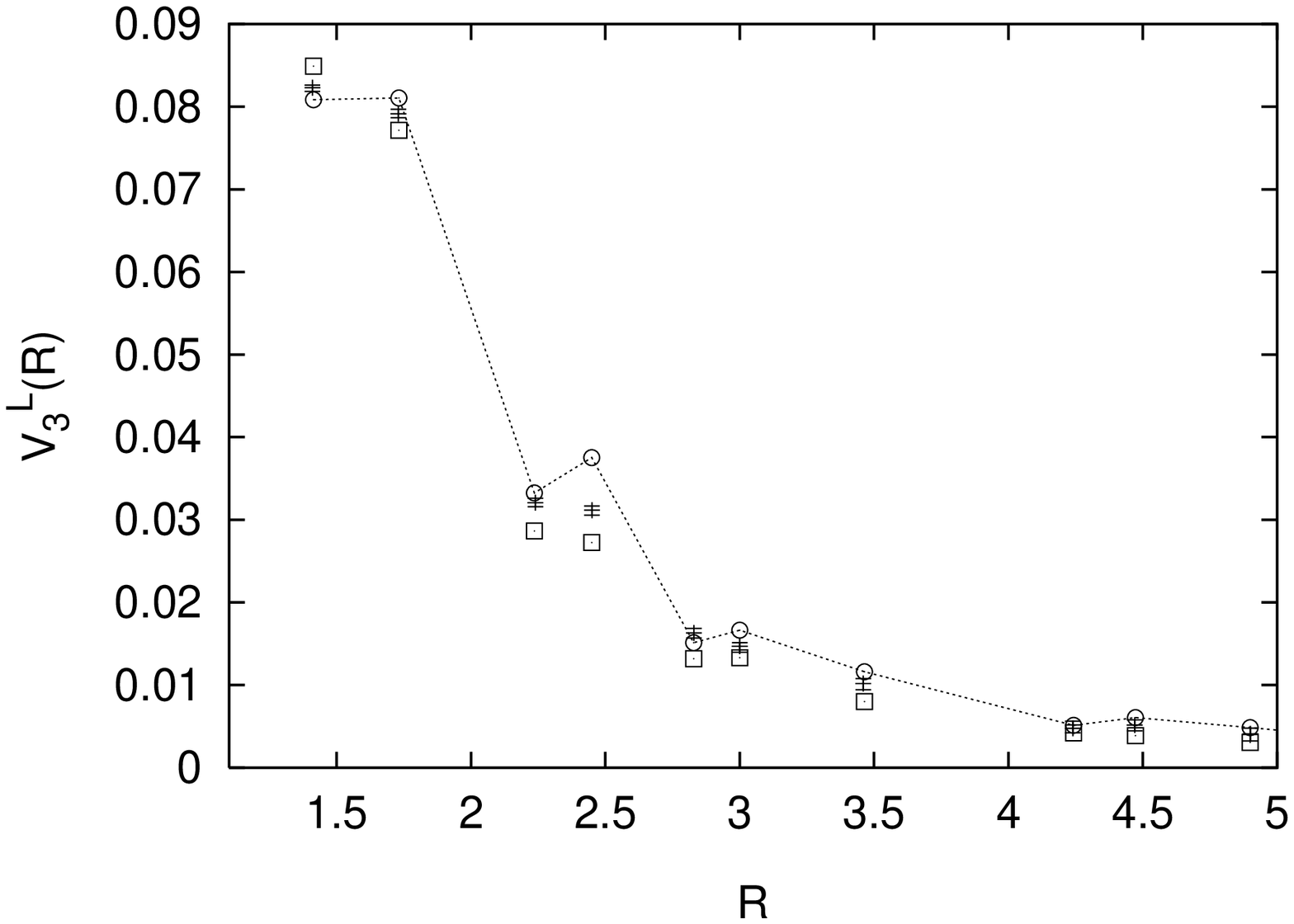}}
\end{picture}
\end{center}
\caption{Same as Fig.~\ref{latt_v2} for $\tilde{V}_3$.}
\label{latt_v3}
\end{figure}

\begin{figure}
\unitlength 1cm
\begin{center}
\begin{picture}(11,8)
\put(0,0){\epsfysize=8cm\epsfbox{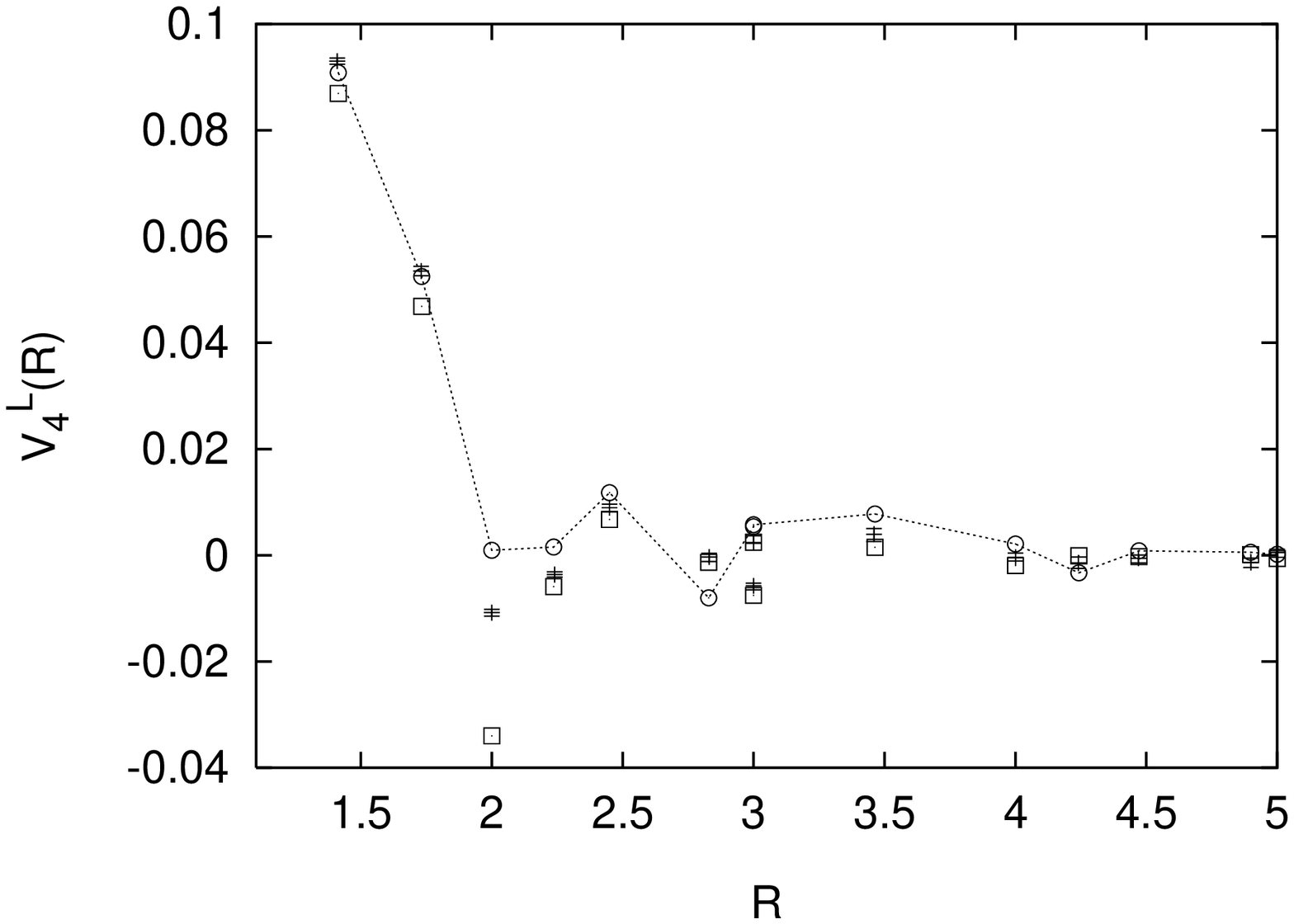}}
\end{picture}
\end{center}
\caption{Same as Fig.~\ref{latt_v2} for $\tilde{V}_4$.}
\label{latt_v4}
\end{figure}

\end{document}